\documentclass[journal]{IEEEtran}
\usepackage{amsmath} 
\makeatletter
\newcommand{\removelatexerror}{\let\@latex@error\@gobble}
\makeatother
\usepackage{cite}
\usepackage{amsfonts,amssymb}
\usepackage{mathtools}
\usepackage{booktabs}
\usepackage{color}
\newtheorem{proposition}{\rm{\textbf{Proposition}}}
\newtheorem{theorem}{\rm{\textbf{Theorem}}}
\newtheorem{assumption}{Assumption}
\newtheorem{definition}{Definition}
\newtheorem{remark}{\rm{\textbf{Remark}}}
\usepackage{multirow}
\usepackage{makecell}
\usepackage{graphicx}
\usepackage{caption}
\usepackage{subcaption}
\usepackage{stfloats}
\begin{document}
\title{Generalizable Learning for Frequency-Domain Channel Extrapolation under Distribution Shift}

\author{Haoyu Wang,~\IEEEmembership{Graduate Student Member,~IEEE,}~Zhi Sun,~\IEEEmembership{Senior Member,~IEEE,} \\
Shuangfeng Han,~\IEEEmembership{Senior Member,~IEEE,}~Xiaoyun Wang,~and Zhaocheng Wang,~\IEEEmembership{Fellow,~IEEE}

\vspace{-20pt}
    \thanks{Haoyu Wang, Zhi Sun, and Zhaocheng Wang are with the Department of Electronic Engineering, Tsinghua University, Beijing 100084 China (e-mail: wanghy22@mails.tsinghua.edu.cn; zhisun@ieee.org; zcwang@tsinghua.edu.cn).}
    \thanks{Shuangfeng Han and Xiaoyun Wang are with the China Mobile Research Institute, Beijing 100053, China. (e-mail: hanshuangfeng@chinamobile.com; wangxiaoyun@chinamobile.com)}
    \thanks{This work was supported by the National Key R\&D Program of China under Grant 2022YFB2902004.}
    \thanks{Corresponding Author: Zhi Sun.}
}

\maketitle

\vspace{-10pt}
\begin{abstract}
   Frequency-domain channel extrapolation is effective in reducing pilot overhead for massive multiple-input multiple-output (MIMO) systems. Recently, Deep learning (DL) based channel extrapolator has become a promising candidate for modeling complex frequency-domain dependency. Nevertheless, current DL extrapolators fail to operate in unseen environments under distribution shift, which poses challenges for large-scale deployment. In this paper, environment generalizable learning for channel extrapolation is achieved by realizing distribution alignment from a physics perspective. Firstly, the distribution shift of wireless channels is rigorously analyzed, which comprises the distribution shift of multipath structure and single-path response. Secondly, a physics-based progressive distribution alignment strategy is proposed to address the distribution shift, which includes successive path-oriented design and path alignment. Path-oriented DL extrapolator decomposes multipath channel extrapolation into parallel extrapolations of the extracted path, which can mitigate the distribution shift of multipath structure. Path alignment is proposed to address the distribution shift of single-path response in path-oriented DL extrapolators, which eventually enables generalizable learning for channel extrapolation. In the simulation, distinct wireless environments are generated using the precise ray-tracing tool. Based on extensive evaluations, the proposed path-oriented DL extrapolator with path alignment can reduce extrapolation error by more than 6 dB in unseen environments compared to the state-of-the-arts.
\end{abstract}

\IEEEpeerreviewmaketitle

\section{Introduction}
Ultra-high spectral efficiency (SE) is a key indicator for B5G and 6G wireless communications, which is essential to support numerous advanced applications such as immersive multi-sensory media \cite{commag_Chukhno_2022_interplay}, and autonomous driving \cite{proceedings_Noor_2022_6G}. Massive multiple-input multiple-output (MIMO) serves as a competitive ultra-high SE solution by leveraging the large-scale antenna array \cite{jstsp_lu_2014_overview}. Despite the large spatial diversity and multiplexing gain, pilot overhead for channel state information (CSI) acquisition also increases with the number of antennas in the antenna array, which limits the effective throughput of massive MIMO systems \cite{twc_wu_2024_environment}. Therefore, low-overhead CSI acquisition is key to enabling future superior services with ultra-high SE in massive MIMO systems. 

Channel extrapolation in the frequency-domain serves as an effective approach to reducing pilot overhead in massive MIMO systems \cite{sigcom_Deepak_2016_eliminating, twc_Rottenberg_2020_performance, tcom_yang_2020_transfer}. Due to the angular and delay reciprocity \cite{jstsp_Han_2019_tracking}, CSI in the different frequency bands shares the same propagation paths in the physical channel, which exhibits strong dependency. As a result, channel extrapolation in frequency-domain aims to infer the CSI in the target frequency band based on measured CSI on another frequency band. For instance, downlink CSI can be extrapolated from the uplink CSI measurement in frequency division duplexing (FDD) massive MIMO systems \cite{sigcom_Deepak_2016_eliminating, twc_Rottenberg_2020_performance, tcom_yang_2020_transfer}, which can eliminate downlink pilot overhead compared to the existing channel estimation and feedback approaches. 

Conventional channel extrapolation approaches are model-based, which calculate the target CSI based on estimated path parameters \cite{sigcom_Deepak_2016_eliminating, twc_Rottenberg_2020_performance}. Typical path parameters estimation algorithms include the interior-point algorithm \cite{sigcom_Deepak_2016_eliminating} and the space-alternating generalized expectation-maximization (SAGE) algorithm \cite{twc_Rottenberg_2020_performance}. However, current path parameters estimation algorithms cannot precisely recover the parameters of each physical path due to the sensitivity to hyper-parameter setting, improper algorithm initialization, and non-convexity in the objective function \cite{sigcom_Deepak_2016_eliminating}. Moreover, the estimation error of path parameters increases due to noise and the interference between unresolvable multipath components. As a result, accurate channel extrapolation cannot be guaranteed when directly calculating the target CSI with the estimated path parameters. 

Deep learning (DL) is an effective approach for accurate channel extrapolation compared to conventional model-based approaches. Thanks to the universal approximation capability of deep neural networks (DNNs) \cite{NN_HORNIK_approximation_1991}, DL extrapolators exhibit advantages in fitting the non-linear dependency of CSI in different frequency bands \cite{Asilomar_Alrabeiah_2019_deep,tcom_yang_2020_transfer,wcl_Yao_2024_loss,mobicom_Liu_2021_FIRE}. Recently, numerous types of DNNs are applied to for channel extrapolation, including multi-layer perceptron (MLP) \cite{Asilomar_Alrabeiah_2019_deep,tcom_yang_2020_transfer}, long short-term memory (LSTM) \cite{wcl_Yao_2024_loss,openj_jiang_2020_deep}, variational autoencoder (VAE) \cite{mobicom_Liu_2021_FIRE}. Despite different DNN structures, most current DL extrapolators operate in an end-to-end manner and can be categorized as channel-oriented, where the input and output of DNNs are CSI composing multiple paths. 

Notwithstanding the great potential in accurate channel extrapolation, DL extrapolators encounter the challenge of environmental generalizability under distribution shift. In the current works of DL extrapolators, training and test datasets are assumed to be drawn from the same distribution. Under the supervised learning framework, DL extrapolators can precisely fit the frequency-domain dependency for the in-distribution CSI samples. However, their extrapolation cannot be guaranteed when the DL extrapolators are tested with out-of-distribution (OOD) channel samples \cite{tkde_wang_2023_generalizing,comst_Akrout_2023_domain,tifs_Rajendran_2022_RF}. Practically, the training dataset for the DL extrapolator is collected within a certain wireless environment, e.g., a street block. Then, the distribution of training CSI samples is determined by the factors that affect the multipath propagation, such as the user positions and building layouts. Since the user positions and building layouts are diverse across different environments, the distribution of CSI samples obviously shifts as well. As a result, OOD samples are inevitable during the operations of the DL extrapolator in a new environment, which leads to the challenge of environment generalizability. 

To enhance the performance of DNNs in the target environments, several learning paradigms have been designed, including domain adaptation \cite{tmc_yin_2024_fewsense}, transfer learning \cite{tcom_yang_2020_transfer,twc_yuan_2021_transfer}, and meta-learning \cite{tcom_yang_2020_transfer,twc_yuan_2021_transfer}. Though environmental generalizability has been brought by the aforementioned learning paradigms, samples from the target environment are still required in their training stages. Explicitly, the DNNs in the training stages of these learning paradigms are firstly pre-trained with datasets from source environments and then fine-tuned with samples from the target environment. However, the time-varying nature of the wireless channel will impose a huge data collection burden when applying the aforementioned learning paradigms. Meanwhile, complex back-propagation and optimization are involved in the fine-tuning stages, which are time-consuming and not suitable for real-time deployment \cite{comst_Akrout_2023_domain}. As a result, the aforementioned learning paradigms cannot achieve generalizable learning for channel extrapolation across environments. 

Environment generalizability without training samples from the target environment can greatly reduce the cost of DL extrapolators in large-scale deployment, which is of practical significance. The current state-of-the-arts for environment generalizability enhancement includes data augmentation \cite{tkde_wang_2023_generalizing,comst_Akrout_2023_domain},  
and hierarchical neural network structure design \cite{mobicom_Banerjee_2024_HORCRUX}. Data augmentation is a widely-used approach to enhance the environmental generalizability of DL-based CSI acquisition, especially for channel feedback \cite{twc_liu_2024_deep,chinacom_han_2024_AI,jstsp_guo_2022_user}. Data augmentation can increase the diversity of CSI samples in the training dataset, which can help the OOD fitting capabilities of DNNs. Typical data augmentation approaches include random circular shift in the angular-delay domain \cite{twc_liu_2024_deep}, flipping \cite{chinacom_han_2024_AI}, and random phase shift \cite{jstsp_guo_2022_user}. Though generalizability gain has been brought empirically, data augmentation still cannot address the curse of environmental generalization. Explicitly, multipath propagation factors that determine the distribution of CSI samples are diverse and complicated. Therefore, the CSI dataset under the multipath propagation conditions of the agnostic target environment cannot be generated by data augmentation schemes. As a result, conventional data augmentation schemes cannot mitigate distribution shifts and will even introduce new distribution biases \cite{cmptsurv_Bayer_2022_survey}. Recently, the hierarchical HORCRUX extrapolator has been proposed \cite{mobicom_Banerjee_2024_HORCRUX}, which is composed of two cascaded layers of neural network channel dividers and mini-neural network distance estimators. Owing to the hierarchical design, HORCRUX extrapolators can generate fine-grained initial guesses for the following path length optimization, which is found to enhance generalizability across environments. However, the neural network channel dividers take the observed channels as input, and the distribution shift is not addressed for the neural networks in different layers of the HORCRUX extrapolators. Consequently, error-prone channel division and initial guesses will occur in the HORCRUX extrapolators under the distribution shift of wireless channels, which will deteriorate the path length optimization and lead to large extrapolation errors. 

In this paper, a physics-based distribution alignment strategy is proposed to enable generalizable learning for channel extrapolation under distribution shift. Firstly, we rigorously model the distribution shift of wireless channels across diverse environments, which is composed of the distribution shift of multipath structure and single-path response. Secondly, a physics-based distribution alignment strategy is proposed to progressively address the distribution shift of wireless channels, which contains two successive steps of path-oriented design and path alignment. Inspired by our previous work \cite{wang2025pathevolutionmodelendogenous}, path-oriented DL extrapolator can decompose the extrapolation of a multipath channel into the extrapolation of each extracted path, which intuitively tackles the distribution shift of multipath structure. Thanks to the path-oriented design, the distribution shift of single-path response can be effectively addressed by path alignment. Thirdly, model training and evaluation are designed to facilitate the path-oriented DL extrapolator and path alignment, including the label co-transformation in the training and output co-compensation in the test stage. Fourthly, the generalization error upper bound of path-oriented with path alignment is derived to theoretically justify the distribution alignment strategy for environmental generalizability enhancement. To evaluate the performance of the proposed path-oriented DL extrapolator, distinct environments are generated in the precise ray-tracing tool Wireless Insite \cite{wi}. Compared to the current state-of-the-arts, path-oriented DL extrapolator with path alignment can reduce extrapolation error for more than 6 dB in unseen environments. Meanwhile, the environment generalizability of path-oriented DL extrapolators with path alignment is consistent with the distribution alignment metric, which lends credibility to our design with theoretical interpretation.

The main contributions of this paper can be summarized as follows.
\begin{itemize}
    \item The distribution shift of wireless channels across diverse environments is rigorously analyzed, which is composed of the distribution shift of multipath structure and single-path response. 
    \item Path-oriented DL extrapolator with path alignment is proposed, which progressively achieves the goal of distribution alignment from a physics-based perspective. 
    \item Model training and evaluation are designed to facilitate the path-oriented DL extrapolator with path alignment. Meanwhile, the effectiveness of the physics-based distribution alignment to the environmental generalizability is theoretically justified. 
\end{itemize}

\textit{Notations}: Uppercase boldface letters denote matrices (e.g., $\mathbf{X}$), while lowercase boldface letters denote vectors (e.g., $\mathbf{x}$). $\mathbb{C}^{m\times n}$ denotes complex spaces with dimension $m\times n$ and ${\rm j}=\sqrt{-1}$. $(\cdot)^{T}$, $(\cdot)^{H}$ denote the operations of transpose and Hermitian transpose. $\text{conj}(\cdot)$ stands for the element-wise conjugate operation. $\otimes$ and $\odot$ denote the Kronecker product and Hadamard product, respectively. $\mathbf{1}_{n}$ denotes an $n$-dimensional column vector with all entries equal to 1. $\mathbf{F}_{n}$ denotes the normalized discrete Fourier transformation (DFT) matrix of size $n\times n$. $\mathbb{E}\{\cdot\}$ denotes the statistical expectation. $U(a,b)$ stands for the uniform distribution between $a$ and $b$. $\delta(x)$ denotes the Dirac function. $[\mathbf{x}]_{n}$ denotes the $n$th element in vector $\mathbf{x}$. $|x|$ denotes the absolute value of scalar $x$. $\Vert\mathbf{x}\Vert$ denotes the Euclidean norm of vector $\mathbf{x}$. $\Vert\mathbf{X}\Vert_{F}$ denotes the Frobenius norm of matrix $\mathbf{X}$. $ \text{supp}(P)$ denotes the support of distribution $P$. 
$\sup$ and $\inf$ denote the supremum and infimum, respectively. $f\circ g$ denotes the composition of functions $f$ and $g$. 

\section{Distribution Shift of Wireless Channel}
In this section, we first present the wideband channel model and frequency-domain channel extrapolation in Sec.~\ref{subsec: model}. The distribution shift of wireless channels is analyzed in Sec.~\ref{subsec: shift analysis}, which causes the environmental generalization challenges of DL extrapolators.
\subsection{Channel Model and Channel Extrapolation}
\label{subsec: model}
In this paper, we consider a wideband massive MIMO system, where a base station (BS) equipped with an $N_{\rm T}$-antenna uniform planar array (UPA) serves a single-antenna user. Based on the geometric Saleh-Valenzuela channel model \cite{twc_he_2023_Beamspace}, the multipath channel $\mathbf{h}(f)\in\mathbb{C}^{N_{\rm T}\times 1}$ with frequency $f$ between the BS and user can be modeled by
\begin{equation}
    \label{equ: channel model}
    \mathbf{h}(f)=\sum_{l=1}^{L}\alpha_{l}e^{-{\rm j}2\pi f\tau_{l}}\mathbf{a}(\varphi_{l}, \theta_{l}), 
\end{equation}
where $L$ denotes the number of paths. For the $l$th path, $\alpha_{l}$ denotes the complex gain, $\tau_{l}$ denotes the delay, $\varphi_{l}$ and $\theta_{l}$ denote the azimuth and elevation angle of arrival (AoA), $\mathbf{a}(\cdot,\cdot)$ denotes the array response. Assume $N_{\rm h}$ and $N_{\rm v}$ antennas are in the horizontal and vertical directions of the UPA. Then, the array response $\mathbf{a}(\varphi,\theta)$ can be calculated by $\mathbf{a}(\varphi,\theta)=\mathbf{a}^{(\rm h)}(\varphi,\theta)\otimes\mathbf{a}^{(\rm v)}(\theta)$ with
\begin{equation}
\label{equ: horizontal array response}
    \begin{aligned}
    \mathbf{a}^{(\rm h)}(\varphi,\theta)&=\frac{1}{\sqrt{N_{\rm h}}}\left[1,e^{{\rm j}2\pi\delta_{\rm h}},\ldots,e^{{\rm j}2\pi(N_{\rm h}-1)\delta_{\rm h}}\right]^{T},\\
    \mathbf{a}^{(\rm v)}(\theta)&=\frac{1}{\sqrt{N_{\rm v}}}\left[1,e^{{\rm j}2\pi\delta_{\rm v}},\ldots,e^{{\rm j}2\pi(N_{\rm v}-1)\delta_{\rm v}}\right]^{T},
    \end{aligned}
\end{equation}
where $\delta_{\rm h}=\frac{d}{\lambda}\sin{(\varphi)}\sin(\theta)$, $\delta_{\rm v}=\frac{d}{\lambda}\cos{(\theta)}$, $d$ denotes the antenna spacing and $\lambda$ denotes the wavelength. Without loss of generality, a half-wavelength spacing with $d=\lambda/2$ is adopted. 

\begin{figure}[t]
        \centering
        \includegraphics[width=0.5\textwidth]{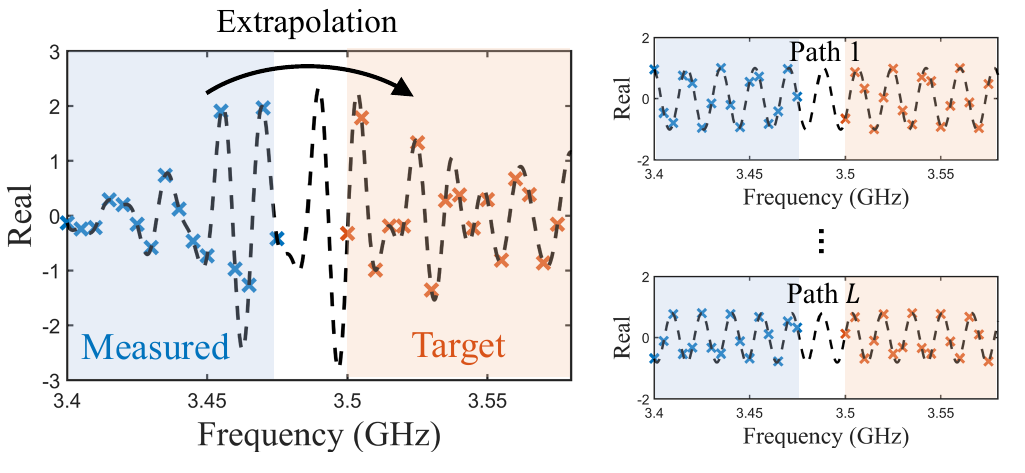}
    \vspace{-10pt}
    \captionsetup{font=footnotesize}
    \caption{Frequency-domain channel extrapolation is illustrated on the left, where the responses of physical paths are shown on the right side. The cross markers denote the measured and target subcarriers. }
    \label{fig: system}
    \vspace{-10pt}
\end{figure}

As shown in Fig.~\ref{fig: system}, channel extrapolation aims to infer the unknown channel in the target frequency band based on the channel in the measured frequency band, which can greatly relieve channel estimation overhead in massive MIMO systems. Denotes the sets of the subcarriers in the measured frequency band and target frequency band by $\mathcal{B}_{\rm m}=\{f_{i}^{(\rm m)}\}_{i=1}^{K_{\rm m}}$ and $\mathcal{B}_{\rm e}=\{f_{i}^{(\rm e)}\}_{i=1}^{K_{\rm e}}$, where $K_{\rm m}$ and $K_{\rm e}$ denotes the number of measured and target subcarriers, respectively. For simplicity, subcarrier frequencies in $\mathcal{B}_{\rm m}$ and $\mathcal{B}_{\rm e}$ are assumed with the same spacing $\Delta f$. With a moderate frequency extrapolation range, spatial reciprocity in the measured and target frequency band is held due to the identical physical paths in the propagation environment \cite{jstsp_Han_2019_tracking}. Based on \eqref{equ: channel model}, the measured channel $\mathbf{H}^{(\rm m)}\in\mathbb{C}^{N_{\rm T}\times K_{\rm m}}$ and target channel $\mathbf{H}^{(\rm e)}\in\mathbb{C}^{N_{\rm T}\times K_{\rm e}}$ can be formulated by 
\begin{equation}
    \label{equ: measured CSI matrix}
    \mathbf{H}^{(x)} = \sum_{l=1}^{L}\alpha_{l}e^{-{\rm j}2\pi f_{1}^{(x)}\tau_{l}}\mathbf{a}(\varphi_{l}, \theta_{l})\mathbf{b}^{H}_{x}(\tau_{l})
\end{equation}
with 
\begin{equation}
    \label{equ: frequency response}
    \mathbf{b}_{x}(\tau)=\left[1,e^{{\rm j}2\pi \Delta f\tau},\ldots,e^{{\rm j}2\pi(K_{x}-1)\Delta f\tau}\right]^{T},
\end{equation}
where $x\in\{\rm m,e\}$ is the indicator for measured or target band. Then, the target channel extrapolation can be formulated as the following mapping
\begin{equation}
    \label{equ: channel extrapolation}
    \psi:\mathbb{C}^{N_{\rm T}\times K_{\rm m}}\to\mathbb{C}^{N_{\rm T}\times K_{\rm e}}, ~ {\mathbf{H}^{(\rm m)}}\mapsto\mathbf{H}^{(\rm e)}=\psi(\mathbf{H}^{(\rm m)}).
\end{equation}

Thanks to the powerful approximation capability, DL extrapolator is usually adopted to fit the target mapping $\psi$. Most existing DL extrapolators are channel-oriented, which are fed with $\mathbf{H}^{(\rm m)}$ as input and try to fit the desired output $\mathbf{H}^{(\rm e)}$ \cite{tcom_yang_2020_transfer,Asilomar_Alrabeiah_2019_deep,mobicom_Liu_2021_FIRE}. Under supervised learning framework, training (source) dataset $\mathcal{D}^{\rm (c)}$ with input and labels can be denoted by 
\begin{equation}
    \label{equ: training dataset}
    \mathcal{D}^{(\rm c)}=\left\{\left(\mathbf{H}_{i}^{(\rm m)}, \mathbf{H}_{i}^{(\rm e)}\right)|i=1,\ldots,M\right\},
\end{equation}
where $M$ denotes the number of training samples, $(\mathbf{H}_{i}^{(\rm m)}, \mathbf{H}_{i}^{(\rm e)})$ denotes the $i$th input-label pair in $\mathcal{D}^{(\rm c)}$. Mean square error (MSE) is adopted as the loss function for the DL extrapolator, which is defined as 
\begin{equation}
    \label{equ: loss definition}
    \mathcal{L}_{G}(\xi)=\mathbb{E}_{(\mathbf{H}^{(\rm m)},\mathbf{H}^{(\rm e)})\sim G}\{\Vert\mathbf{H}^{(\rm e)}-\xi(\mathbf{H}^{(\rm m)})\Vert_{F}^{2}\}
\end{equation}
for samples distribution $G$ and DL extrapolator $\xi$. Then, the trained channel-oriented DL extrapolator $\xi^{*}$ is yielded by minimizing the training loss, i.e., $\xi^{*}=\mathop{\arg\min}_{\xi\in\mathcal{F}}\mathcal{L}_{\mathcal{D}^{(\rm c)}}(\xi)$. Hereby, $\mathcal{F}$ denotes the hypothesis space of the DL extrapolator, which is determined by the adopted neural network type and its structure configurations. 

\subsection{Distribution Shift Analysis}
\label{subsec: shift analysis}
Understanding the distribution shift is vital to enhancing the environmental generalizability of the DL extrapolator. Assume that training dataset $\mathcal{D}^{(\rm c)}$ is drawn from joint distribution $(\mathbf{H}^{(\rm m)},\mathbf{H}^{(\rm e)})\sim P^{(c)}$. According to the Bayes rule, the joint distribution $P^{(\rm c)}$ can be factorized as
\begin{equation}
    \label{equ: conditional}
    P^{(\rm c)}=P_{\mathbf{H}^{(\rm e)}|\mathbf{H}^{(\rm m)}}^{(\rm c)}\times P_{\mathbf{H}^{(\rm m)}}^{(\rm c)}.
\end{equation}
Hereby, the distribution shift includes two parts, namely, shift of conditional distribution $P_{\mathbf{H}^{(\rm e)}|\mathbf{H}^{(\rm m)}}^{(\rm c)}$ and shift of the marginal distribution $P_{\mathbf{H}^{(\rm m)}}^{(\rm c)}$. Based on the existence of the underlying mapping $\psi$, the conditional distribution $P_{\mathbf{H}^{(\rm e)}|\mathbf{H}^{(\rm m)}}^{(\rm c)}$ can be formulated as $P_{\mathbf{H}^{(\rm e)}|\mathbf{H}^{(\rm m)}}^{(\rm c)}=\delta(\mathbf{H}^{(\rm e)}-\psi(\mathbf{H}^{(\rm m)}))$. Then, the shift of $P_{\mathbf{H}^{(\rm e)}|\mathbf{H}^{(\rm m)}}^{(\rm c)}$ is equivalent to the shift of underlying mapping $\psi$, which is given as follows.
\begin{proposition}
    \label{prop: sharing xi}
    For a wideband massive MIMO system with sparse channels, an invariant mapping $\psi$ can be assumed among different wireless environments. 
\end{proposition}
\begin{IEEEproof}
With high-resolution in angular and delay domain for wideband massive MIMO system, underlying mapping $\psi_{1}$ that from $\mathbf{H}^{(\rm m)}$ to path parameters $\{\alpha_{l}, \tau_{l}, \varphi_{l}, \theta_{l}\}_{l=1}^{L}$ exists for sparse wireless channels. Owing to the fixed system parameter configurations (i.e., fixed parameters $\{N_{\rm h}, N_{\rm v}, K_{\rm m}, \Delta f\}$) and the universal channel model in \eqref{equ: channel model}, an identical $\psi_{1}$ can be assumed among different environments. Once the path parameters $\{\alpha_{l}, \tau_{l}, \varphi_{l}, \theta_{l}\}_{l=1}^{L}$ is obtained, the target channel $\mathbf{H}^{(\rm e)}$ is yielded based on \eqref{equ: measured CSI matrix}, which can be denoted by the mapping $\psi_{2}$. Similarly, with fixed system setting of $\{N_{\rm h}, N_{\rm v}, K_{\rm e}, \Delta f\}$, an invariant $\psi_{2}$ among different environments can be obtained as well. Since $\psi=\psi_{2}\circ\psi_{1}$ is held, common $\psi$ among different environments can be achieved. 
\end{IEEEproof}

Shift of the marginal distribution $P_{\mathbf{H}^{(\rm m)}}^{(\rm c)}$ is the main cause to the shift of joint distribution $P^{(\rm c)}$. As a result of the multipath effect, $\mathbf{H}^{(\rm m)}$ is the sum of the response of multiple paths, which complicates $P_{\mathbf{H}^{(\rm m)}}^{(\rm c)}$. Hereby, we define the response of $l$th physical path at the measured frequency band as
\begin{equation}
    \label{equ: path channel}
    \mathbf{A}_{l}=\alpha_{l}e^{-{\rm j}2\pi f_{1}^{(\rm m)}\tau_{l}}\mathbf{a}(\varphi_{l}, \theta_{l})\mathbf{b}^{H}_{\rm m}(\tau_{l}),
\end{equation}
and the measured channel can be reformulated as $\mathbf{H}^{(\rm m)}=\sum_{l=1}^{L}\mathbf{A}_{l}$. With the law of total probability, the probability density of $\mathbf{H}^{(\rm m)}$ can be factorized as 
\begin{equation}
    \label{equ: marginal expansion}
    \begin{aligned}
    &p(\mathbf{H}^{(\rm m)})=\mathbb{E}_{L}\{p(\mathbf{H}^{(\rm m)}|L)\}\\
    &=\sum_{k}\underbrace{P(L=k)}_{\text{factor (i)}}\int_{\Omega_{L}} \underbrace{p(\mathbf{A}_{1},\ldots,\mathbf{A}_{L}|L=k)}_{\text{factor (ii) and (iii)}}\prod_{l=1}^{L} d\mathbf{A}_{l},
    \end{aligned}
\end{equation}
where $P(L=k)$ denotes the probability when the number of paths equals $k$, $p(\mathbf{A}_{1},\ldots,\mathbf{A}_{L}|L=k)$ denotes the joint distribution of $(\mathbf{A}_{1},\ldots,\mathbf{A}_{L})$ and $\Omega_{L}=\{(\mathbf{A}_{1},\ldots,\mathbf{A}_{L})|\sum_{l=1}^{L}\mathbf{A}_{l}=\mathbf{H}^{(\rm m)}\}$. 
\begin{figure}[t]
    \centering
    \includegraphics[width=0.45\textwidth]{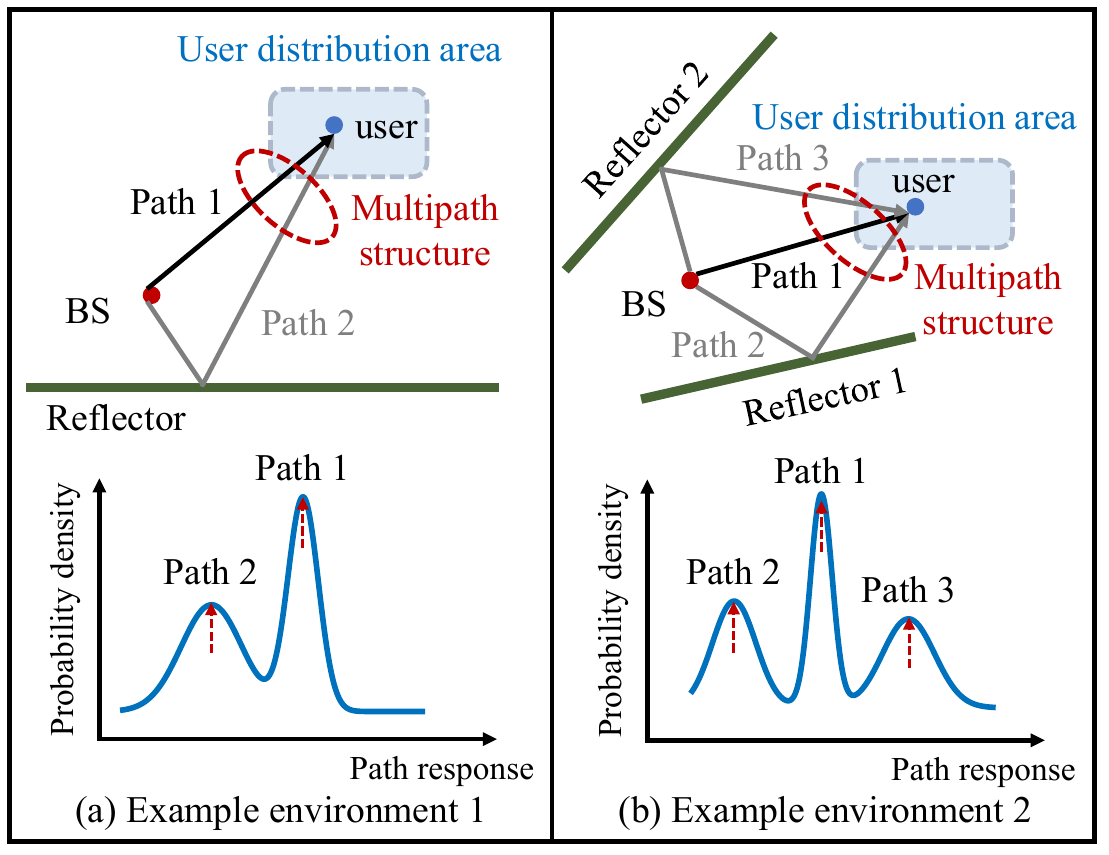}
    \captionsetup{font=footnotesize}
    \caption{Illustration of distribution shift between two example environments. The 2D geometrical layouts of two environments are shown on the top, where users are distributed within the blue box. Accordingly, the marginal distributions of path response in two environments are shown at the bottom.}
    \label{fig: shift model}
    \vspace{-10pt}
\end{figure}
As shown in the underbraces in \eqref{equ: marginal expansion}, the marginal distribution $p(\mathbf{H}^{(\rm m)})$ is impacted by 3 different factors: (i) distribution of the number of paths; (ii) multipath dependency; (iii) marginal distribution of single-path response. For expression simplicity, factors (i) and (ii) can be merged as the distribution of multipath structure.

Two example environments are considered to illustrate the distribution shift of the wireless channels, which are shown in Fig.~\ref{fig: shift model}. Due to the variations in the number and deployment of reflectors, there are two paths in the example environment 1, while three paths in the example environment 2. Additionally, the paths in the two environments exhibit different relative relationships in geometrical and electromagnetic properties, leading to the shift of multipath dependency. Both varying numbers of paths and multipath dependencies lead to the distribution shift of the multipath structure. Moreover, considering the distinctions of wireless propagation conditions and user position distributions, the marginal distributions of single-path response obviously vary as well, which are shown at the bottom of Fig.~
\ref{fig: shift model}. As a result, the distinctions in propagation conditions and user positions lead to the distribution shift of wireless channels across environments. 

The distribution shift of wireless channels between the training and test environments leads to the environment generalization problem under the supervised learning framework. By minimizing training loss $\mathcal{L}_{\mathcal{D}^{(\rm c)}}(\xi)$, trained DL extrapolator $\xi^{*}$ can achieve high approximation accuracy for samples $\mathbf{H}^{(\rm m)}$ in distribution $P_{\mathbf{H}^{(\rm m)}}^{(\rm c)}$ from the source environment. However, for the samples out of $P_{\mathbf{H}^{(\rm m)}}^{(\rm c)}$, their fitting accuracy cannot be guaranteed with the trained DL extrapolator $\xi^{*}$. Considering the aforementioned distribution shift of $P_{\mathbf{H}^{(\rm m)}}^{(\rm c)}$, the performance of the trained extrapolator will obviously degrade in the new environments with distinct propagation conditions, which poses challenges to the deployment of DL extrapolators. 

\begin{remark}
\label{remk: distribution shift}
Distribution shift of the wireless channel is composed of the distribution shift of multipath structure and single-path response, which are led by diverse propagation-related factors (e.g., building layouts and user positions) across environments. Consequently, numerous OOD channel samples can occur in unseen test environments, which drastically degrades the performance of trained DL extrapolators. 
\end{remark}

\section{Reducing Distribution Shift by Path-Oriented DL Extrapolator with Path Alignment}
In Sec.~\ref{subsec: strategy}, a physics-based progressive distribution alignment strategy is proposed to address the distribution shift of wireless channels, which contains path-oriented DL extrapolator design and path alignment. Then, path-oriented DL extrapolator design and path alignment are detailed in Sec.~\ref{subsec: pe formulation} and Sec.~\ref{subsec: alignment}.

\subsection{Physics-Based Progressive Distribution Alignment}
\label{subsec: strategy}
\begin{figure}[t]
        \centering
        \includegraphics[width=0.48\textwidth]{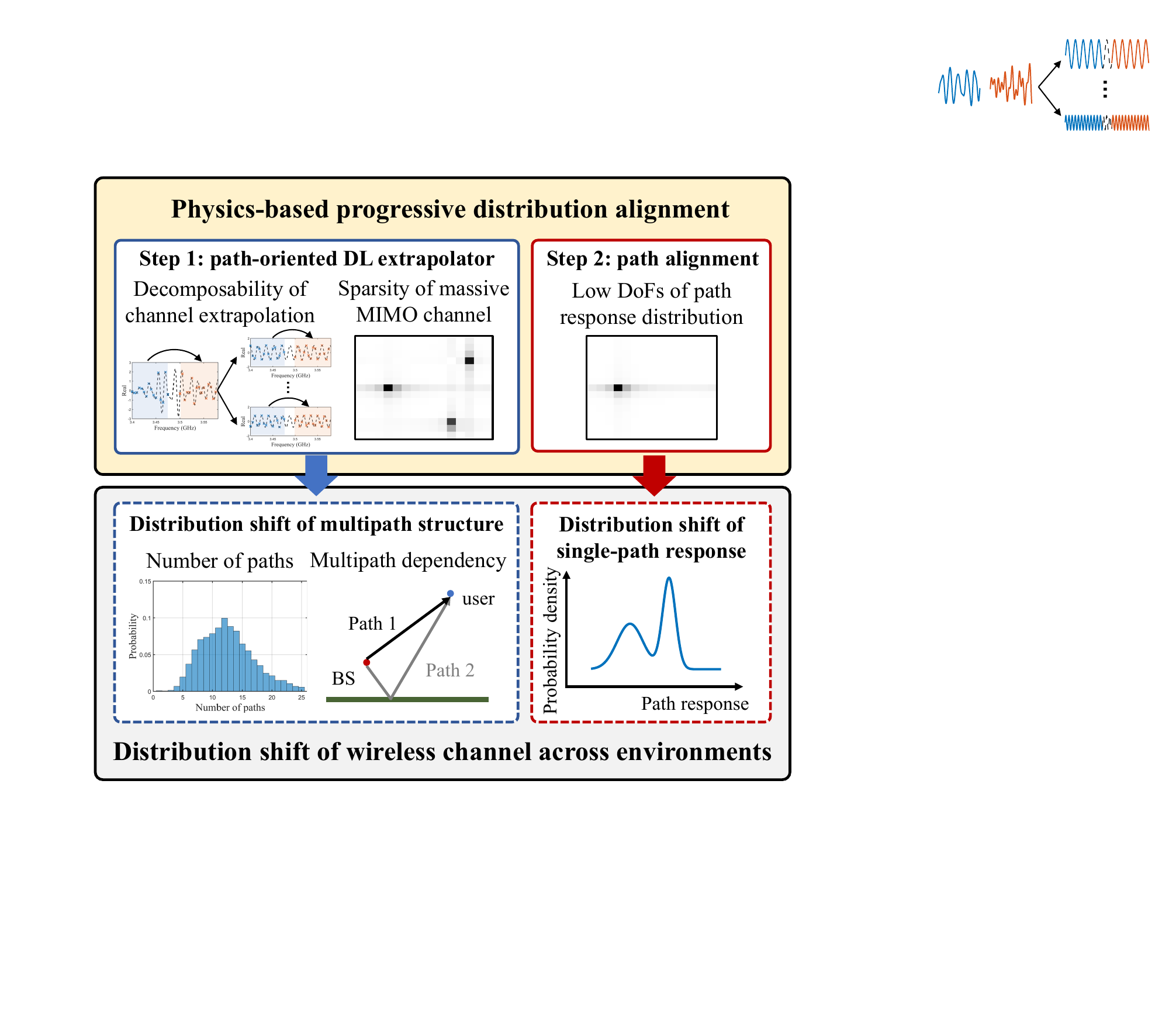}
    \captionsetup{font=footnotesize}
    \caption{Illustration of physics-based distribution alignment strategy, which can progressively reduce the distribution shift of wireless channel across environments.}
    \label{fig: reduction}
    \vspace{-10pt}
\end{figure}
Input distribution alignment can fundamentally enhance the environment generalizability of DL extrapolators. Under the supervised learning framework, the neural network can approximate the target function $\psi$ under the distribution of input in the training dataset. Based on the environment-invariance of the target function in \textbf{Proposition}~\ref{prop: sharing xi}, the generalizability of the trained neural network can be guaranteed when the input distribution in the test environment is aligned with the training environment. To this end, a physics-based distribution alignment is proposed based on the distribution shift analysis in Sec.~\ref{subsec: shift analysis}, which is discussed as follows.

Path-oriented DL extrapolator can effectively mitigate the distribution shift of multipath structure, which is illustrated on the left of Fig.~\ref{fig: reduction}. From the perspective of physics, the extrapolation of multipath channels is path-wise decomposable. Explicitly, the extrapolation of the multipath channel can be decomposed into the extrapolation of each path, i.e., 
\begin{equation}
    \label{equ: decompse}
    \mathbf{H}^{(\rm e)}=\psi(\mathbf{H}^{(\rm m)})=\sum_{l=1}^{L}\psi(\mathbf{A}_{l}).
\end{equation}
Meanwhile, the wireless channel exhibits sparsity in the angular-delay domain in a wideband massive MIMO system, which facilitates the extraction of path response from the measured multipath channel. Based on the aforementioned decomposability and sparsity, we propose the path-oriented DL extrapolator design, which first extracts path response and then parallelly extrapolates each extracted path via DL. For the trained path-oriented DL extrapolator, the neural network exclusively captures the frequency-domain correlations of each extracted path response, which decouples the number and the dependency of the extracted paths. Therefore, the shift of multipath structure can be effectively mitigated.

Path alignment can reduce the distribution shift of single-path response, which is illustrated in the right of Fig.~\ref{fig: reduction}. Compared to the distribution of multipath channel, the distribution of extracted path response has relatively low degrees of freedom (DoFs), which facilitates the design of path alignment approach.

\begin{remark}
    \label{remk: alignment}
    Distribution alignment of wireless channels is achieved in a logically progressive manner. Firstly, path-oriented design removes the distribution shift of the multipath structure. Then, path alignment can further reduce the distribution shift of extracted paths, which eventually enables distribution alignment.
\end{remark}
\begin{remark}
    \label{remk: knowledge function}
    Physics-based design is necessary for distribution alignment. Explicitly, path extraction and path alignment are both designed based on the knowledge of channel model, which is free of any training data. On the contrary, although a divide-and-conquer design is proposed in the hierarchical HORCRUX extrapolator to enhance environmental generalizability, the divisions of subchannels are also data-driven \cite{mobicom_Banerjee_2024_HORCRUX}. Meanwhile, the marginal distribution shift in each subchannel is not addressed, either. Consequently, the severe distribution shift of training and agnostic test channel samples will degrade the environmental generalizability of the HORCRUX extrapolator.
\end{remark}
\subsection{Step 1: Path-Oriented DL Extrapolator Design}
\label{subsec: pe formulation}
\begin{figure}[t]
        \centering
        \includegraphics[width=0.48\textwidth]{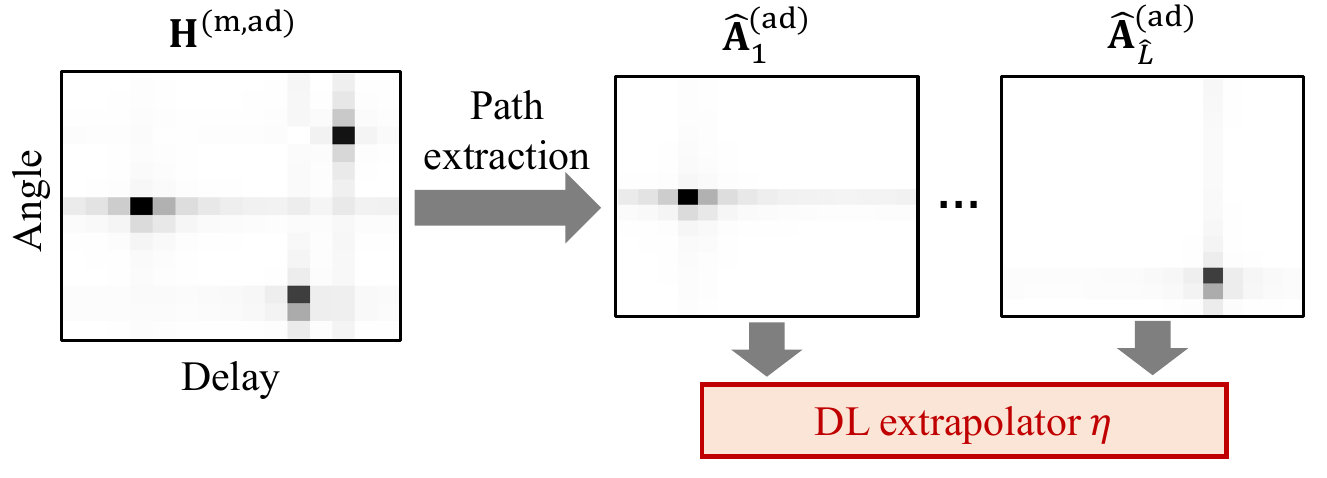}
    \captionsetup{font=footnotesize}
    \caption{Illustration of path-oriented DL extrapolator design. Multipath channel and extracted path responses at the measured frequency band are shown in the angular-delay domain, where darker colors represent stronger power.}
    \label{fig: extraction}
    \vspace{-10pt}
\end{figure}
Path-oriented DL extrapolator is the first step for distribution alignment by mitigating the distribution shift of the multipath structure. An overview of the path-oriented DL extrapolator is illustrated in Fig.~\ref{fig: extraction}. Explicitly, we first extract path response $\{\widehat{\mathbf{A}}_{l}\}_{l=1}^{\hat{L}}$ from $\mathbf{H}^{(\rm m)}$, where $\hat{L}$ denotes the number of extracted paths. Denote the angular-delay representations by $\mathbf{H}^{(\rm{m,ad})}=\mathbf{T}\mathbf{H}^{(\rm m)}\mathbf{F}_{K_{\rm m}}^{H}$ and $\widehat{\mathbf{A}}^{(\rm{ad})}_{l}=\mathbf{T}\widehat{\mathbf{A}}^{(\rm m)}_{l}\mathbf{F}_{K_{\rm m}}^{H}$ with $\mathbf{T}=\mathbf{F}_{N_{\rm h}}\otimes\mathbf{F}_{N_{\rm v}}$. As shown in Fig.~\ref{fig: extraction}, the power of the extracted path response is concentrated in the angular-delay domain. To this end, we employ the SAGE \cite{jsac_Fleury_1999_sage} and clustering algorithms to extract path responses. Each extracted response aims to approximate the response of an individual physical path or the aggregate responses of multiple physical paths exhibiting similar angle and delay parameters. Firstly, SAGE algorithm is applied to $\mathbf{H}^{(\rm m)}$ to output the estimated parameters $\widehat{\Theta}^{(\rm m)}=\{\hat{\alpha}_{k}^{(\rm m)},\hat{\varphi}_{k}^{(\rm m)},\hat{\theta}_{k}^{(\rm m)},\hat{\tau}_{k}^{(\rm m)}\}_{k=1}^{\hat{L}^{(\rm m)}}$. Due to the alternating nature and limited precision in the SAGE algorithm, the estimated parameters in $\widehat{\Theta}^{(\rm m)}$ are not the ground-truth parameters $\{\alpha_{l},\varphi_{l},\theta_{l},\tau_{l}\}_{l=1}^{L}$ of physical paths. Practically, the estimated parameters in $\widehat{\Theta}^{(\rm m)}$ usually emerge as multiple sub-paths surrounding the ground-truth path parameters, which can be found in real-world channel measurement \cite{tap_Ling_2017_Experimental}. Therefore, we adopt the DBSCAN algorithm \cite{tods_Schubert_2017_dbscan} for clustering, which can automatically choose the optimal number of clusters. To facilitate DBSCAN clustering, the distance $D(\hat{\boldsymbol{\omega}}_{k},\hat{\boldsymbol{\omega}}_{l})$ between the estimated paths is defined as
\begin{equation}
    \label{equ: distance}
    D(\hat{\boldsymbol{\omega}}_{k},\hat{\boldsymbol{\omega}}_{l})=\sqrt{\sum_{i=1}^3|e^{{\rm j}[\hat{\boldsymbol{\omega}}_{k}]_i}-e^{{\rm j}[\hat{\boldsymbol{\omega}}_{l}]_i}|^2},
\end{equation}
where $\hat{\boldsymbol{\omega}}_{k}\!=\!\big[\pi\!\sin{\!(\hat{\varphi}_{k}^{(\rm m)})}\sin{\!(\hat{\theta}_{k}^{(\rm m)})},\pi\!\cos{\!(\hat{\theta}_{k}^{(\rm m)})},2\pi\Delta f\hat{\tau}_{k}^{(\rm m)}\big]^T$. With the clustered parameters, the response $\widehat{\mathbf{A}}_{l}$ of $l$th extracted path can be obtained by summing up the response of each sub-path in $l$th cluster, i.e., 
\begin{equation}
\label{equ: extracted path}
\widehat{\mathbf{A}}_{l}\!=\!\sum_{k\in\mathcal{I}^{(\rm m)}_{l}}\hat{\alpha}_{k}^{(\rm m)}e^{-{\rm j}2\pi f_{1}^{(\rm m)}\hat{\tau}_{k}^{(\rm m)}}\mathbf{a}(\hat{\varphi}_{k}^{(\rm m)}, \hat{\theta}_{k}^{(\rm m)})\mathbf{b}_{\rm m}^{H}(\hat{\tau}_{k}^{(\rm m)}),
\end{equation}
where $\mathcal{I}^{(\rm m)}_{l}$ denotes the index subset of $l$th cluster. Then, we can parallelly extrapolate the extracted path response $\widehat{\mathbf{A}}_{l}$ via DL. 

Different from the model-based channel extrapolator, the proposed path-oriented DL extrapolator directly relies on the precision of the extracted responses $\{\widehat{\mathbf{A}}_{l}\}$, which is more robust to the parameter estimation error. On the one hand, a low approximation error $\Vert\mathbf{H}^{(\rm m)}-\sum_{l=1}^{\hat{L}}\widehat{\mathbf{A}}_{l}\Vert_{F}^2$ can be achieved considering the monotonicity property of SAGE algorithm \cite{jsac_Fleury_1999_sage}. On the other hand, with DBSCAN clustering, $\widehat{\mathbf{A}}_{l}$ can approximate the response of an individual physical path or multiple physical paths with similar angle and delay parameters, which enables robust path extraction. Additionally, since the accurate parameters of the physical paths are still unattainable from $\widehat{\mathbf{A}}_{l}$, especially in the presence of unresolvable physical paths and noise, DL is still required to achieve accurate and robust channel extrapolation. 

\subsection{Step 2: Path Alignment}
\label{subsec: alignment}
\begin{figure}[t]
        \centering
        \includegraphics[width=0.45\textwidth]{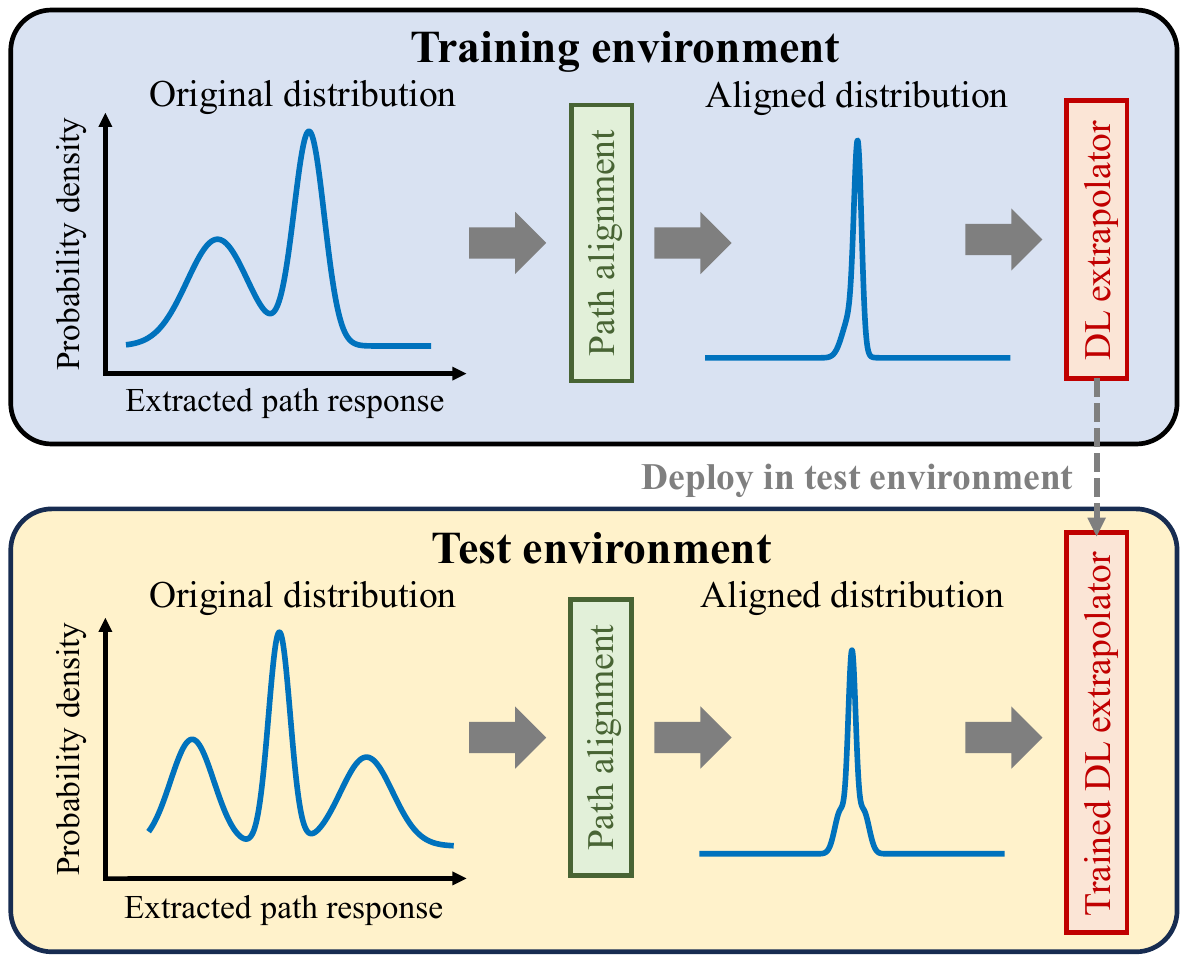}
    \captionsetup{font=footnotesize}
    \caption{Illustration of path alignment to align the distribution of extracted path response between the training and unseen test environment. }
    \label{fig: alignment}
    \vspace{-10pt}
\end{figure}

Building upon the path-oriented design, path alignment is proposed to reduce the distribution shift of the extracted path response. The overview of path alignment is illustrated in Fig.~\ref{fig: alignment}. By applying the path alignment during both training and testing, the distinct distributions of extracted path responses can be aligned with a much lower variance. Thus, when the path-oriented DL extrapolator is trained with the aligned path response, its performance can still be guaranteed when input with the aligned path response in the unseen test environments. 

\begin{figure}[t]
        \centering
        \includegraphics[width=0.45\textwidth]{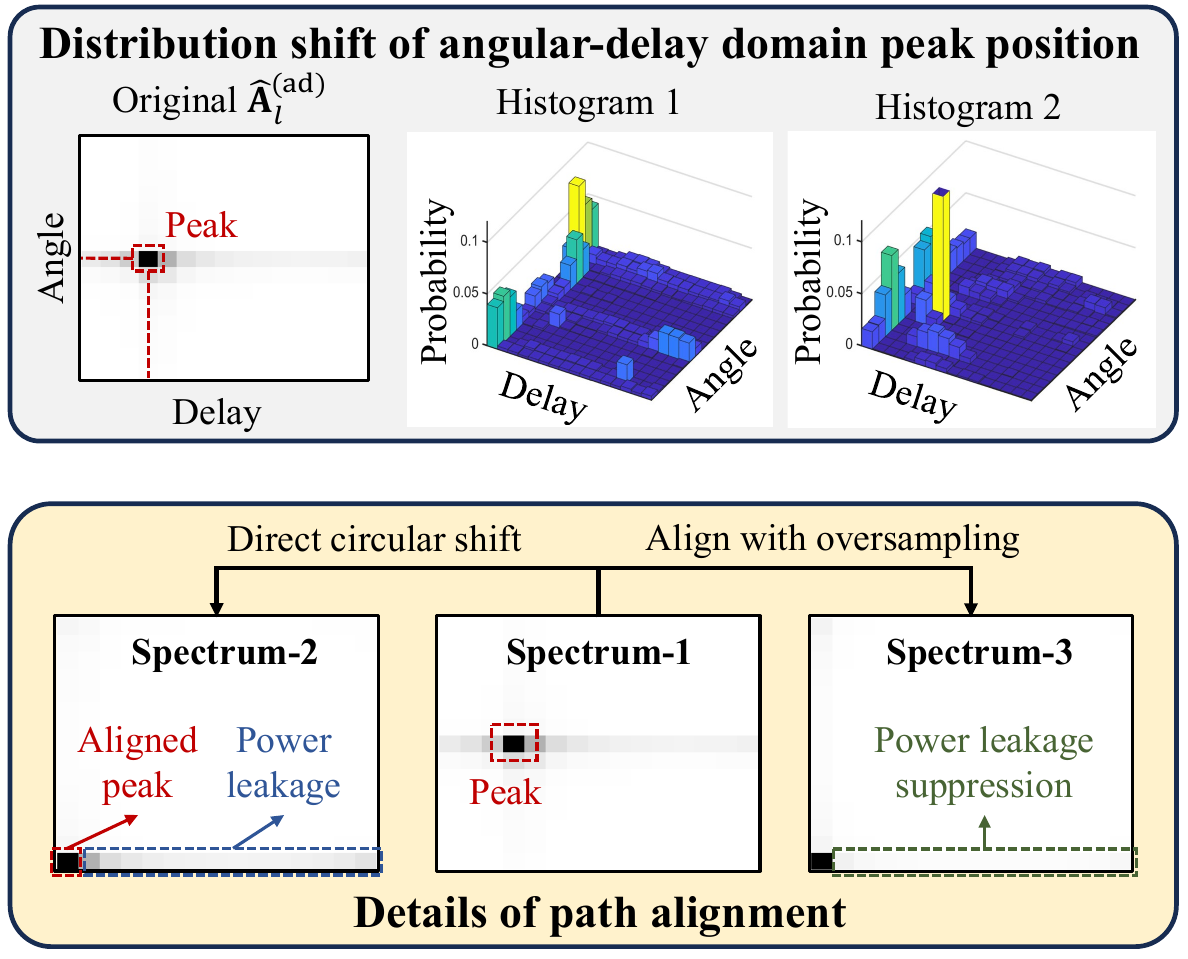}
    \captionsetup{font=footnotesize}
    \caption{Distribution shift of angular-delay domain peak position of extracted path response is shown on the top. Details of path alignment are illustrated at the bottom. 
    }
    \label{fig: peak position shift}
    \vspace{-10pt}
\end{figure}

Understanding the distribution shift of extracted path response is essential for path alignment. Due to the bijective relationship between $\widehat{\mathbf{A}}_{l}$ and its angular-delay domain representation $\widehat{\mathbf{A}}^{(\rm ad)}_{l}$, it is equivalent to analyze the distribution shift of $\widehat{\mathbf{A}}_{l}^{(\rm ad)}$. The original $\widehat{\mathbf{A}}^{(\rm ad)}_{l}$ is shown in the top left of Fig.~\ref{fig: peak position shift}. It can be found that $\widehat{\mathbf{A}}_{l}^{(\rm ad)}$ is sparse in the angular-delay domain with a prominent peak, which dominates the power of the extracted path response. Therefore, the distribution shift of the position and value of the peak element in $\widehat{\mathbf{A}}_{l}^{(\rm ad)}$ plays a critical role in the distribution shift of $\widehat{\mathbf{A}}_{l}^{(\rm da)}$. The impact of peak value shift can be mitigated by normalization. On the contrary, the peak position is related to extracted geometrical parameters, incorporating diverse environmental propagation factors. For instance, peak position histograms 1 and 2 in the top of Fig.~\ref{fig: peak position shift} are yielded from two environments, which exhibit obvious distribution shift. Thus, the distribution shift of the angular-delay domain peak position dominates the distribution of the extracted path response. 

To effectively reduce the peak position distribution shift among environments, a path alignment approach is designed as follows. Based on the property of DFT, circular shift in the rows and columns of $\widehat{\mathbf{A}}_{l}^{(\rm ad)}$ is equivalent to shifting the estimated geometrical parameters $\{\hat{\tau}_{k}^{(\rm m)}, \hat{\varphi}_{k}^{(\rm m)},\hat{\theta}_{k}^{(\rm m)}\}_{k\in\mathcal{I}^{(\rm m)}_{l}}$. Thus, we can circularly shift $\widehat{\mathbf{A}}_{l}^{(\rm ad)}$ to relocate the peak element of $\widehat{\mathbf{A}}_{l}^{(\rm ad)}$ to a fix position, which is depicted as the spectrum-2 in Fig.~\ref{fig: peak position shift}. When the peak position is aligned, the distribution shift of extracted path response between training and test environments can be effectively reduced. However, due to the finite number of antennas and measured subcarriers, power leakage occurs in the angular-delay domain and cannot be mitigated by circular shift, which is illustrated in the blue box of the spectrum-2 in Fig.~\ref{fig: peak position shift}. Owing to the power leakage effect, the power portion of the elements near the peak element increases. As a result, power leakage will limit the distribution alignment performance via direct circular shift. 

To mitigate the power leakage effect, we can scan the peak position with oversampling factors for fine-grained path alignment. Based on the orthogonality of the angular-delay representation and Parseval's Theorem, a peak element with higher power can be searched via oversampling, while the power of the non-peak elements is reduced, which mitigates the power leakage effect. Denote horizontal and vertical oversampling factors as $O_{\rm h}$ and $O_{\rm v}$, the horizontal scanning vector $\mathbf{w}_{n_{1}}^{(\rm h)}~(n_{1}=0,\ldots,O_{\rm h}N_{\rm h}-1)$ and vertical scanning vector $\mathbf{w}_{n_{2}}^{(\rm v)}~(n_{2}=0,\ldots,O_{\rm v}N_{\rm v}-1)$ can be represented by 
\begin{equation}
    \label{equ: angular scan vector}
    \begin{aligned}
    \mathbf{w}_{n_{1}}^{(\rm h)}&=\Big[1,e^{{\rm j}2\pi \frac{n_{1}}{O_{\rm h}N_{\rm h}}},\ldots,e^{{\rm j}2\pi\frac{n_{1}(N_{\rm h}-1)}{O_{\rm h}N_{\rm h}}}\Big]^{T},\\
    \mathbf{w}_{n_{2}}^{(\rm v)}&=\Big[1,e^{{\rm j}2\pi\frac{n_{2}}{O_{\rm v}N_{\rm v}}},\ldots,e^{{\rm j}2\pi\frac{n_{2}(N_{\rm v}-1)}{O_{\rm v}N_{\rm v}}}\Big]^{T}.
    \end{aligned}
\end{equation}
Denote the delay domain oversampling factor as $O_{\rm d}$, the delay domain scanning vector $\mathbf{w}_{n_{3}}^{(\rm m)}~(n_{3}=0,\ldots,O_{\rm d}K_{\rm m}-1)$ for measured frequency band can be represented by
\begin{equation}
    \label{equ: delay scan vector (measured)}
    \mathbf{w}^{(\rm m)}_{n_{3}}=\Big[1,e^{{\rm j}2\pi\frac{n_{3}}{O_{\rm d}K_{\rm m}}},\ldots,e^{{\rm j}2\pi\frac{n_{3}(K_{\rm m}-1)}{O_{\rm d}K_{\rm m}}}\Big]^{T}.
\end{equation}
By scanning with $\mathbf{w}^{(\rm h,v)}_{n_{1},n_{2}}=\mathbf{w}^{(\rm h)}_{n_{1}}\otimes\mathbf{w}^{(\rm v)}_{n_{2}}$, the peak position $(\tilde{n}_{1,l},\tilde{n}_{2,l})$ in angular-domain can be calculated by 
\begin{equation}
    \label{equ: angular peak}
    (\tilde{n}_{1,l},\tilde{n}_{2,l})=\mathop{\arg\max}_{(n_{1},n_{2})}\left\{\Vert(\mathbf{w}_{n_{1},n_{2}}^{(\rm h,v)})^{H}\widehat{\mathbf{A}}_{l}\Vert^{2}_{2}\right\}.
\end{equation}
Similarly, peak position $\tilde{n}_{3,l}$ in the delay-domain can be calculated by
\begin{equation}
    \label{equ: delay peak}
    \tilde{n}_{3,l}=\mathop{\arg\max}_{n_{3}}\left\{\Vert\widehat{\mathbf{A}}_{l}\mathbf{w}_{n_{3}}^{(\rm m)}\Vert^{2}_{2}\right\}.
\end{equation}
Next, the peak position in the angular-delay domain can be equivalently aligned by adjusting the phase of each element in $\widehat{\mathbf{A}}_{l}$. Explicitly, the angular-domain phase adjusting matrix $\mathbf{S}_{\rm a}^{(\rm m)}(n_{1},n_{2})\in\mathbb{C}^{N_{\rm T}\times K_{\rm m}}$ with peak position $(n_{1},n_{2})$ and the delay-domain phase adjusting matrix $\mathbf{S}_{\rm d}^{(\rm m)}(n_{3})\in\mathbb{C}^{N_{\rm T}\times K_{\rm m}}$ with peak position $n_{3}$ can be defined as 
\begin{equation}
    \label{equ: shift matrix in measured band}
    \begin{aligned}
    \mathbf{S}_{\rm a}^{(\rm m)}(n_{1},n_{2})&=\text{conj}\big(\mathbf{w}^{(\rm h,v)}_{n_{1},n_{2}}\big)\otimes\mathbf{1}_{K_{\rm m}}^{T},\\
    \mathbf{S}_{\rm d}^{(\rm m)}(n_{3})&=\mathbf{1}_{N_{\rm T}}\otimes(\mathbf{w}_{n_{3}}^{(\rm m)})^{T}.
    \end{aligned}
\end{equation} 
With scanned peak position $(\tilde{n}_{1,l},\tilde{n}_{2,l},\tilde{n}_{3,l})$, the aligned path response $\widetilde{\mathbf{A}}_{l}$ can be calculated by
\begin{equation}
    \label{equ: aligned feature}
    \widetilde{\mathbf{A}}_{l} = \big(\mathbf{S}_{\rm a}^{(\rm m)}(\tilde{n}_{1,l},\tilde{n}_{2,l})\odot\mathbf{S}_{\rm d}^{(\rm m)}(\tilde{n}_{3,l})\big)\odot\widehat{\mathbf{A}}_{l}\triangleq \mathbf{U}_{l}\odot\widehat{\mathbf{A}}_{l}.
\end{equation}
The angular-delay domain representation of $\widetilde{\mathbf{A}}_{l}$ is depicted in the spectrum-3 of Fig.~\ref{fig: peak position shift}. It can be found that the peak position of spectrum-3 is aligned as well. Meanwhile, power leakage can be effectively suppressed in spectrum-3 compared to spectrum-2, which further reduces the distribution shift.

\begin{remark}
    \label{remk: distinction}
    Proposed path alignment is distinct from conventional data augmentation \cite{twc_liu_2024_deep,jstsp_guo_2022_user,chinacom_han_2024_AI}. The path alignment approach aims to reduce the distribution shift of wireless channels, which deterministically process the extracted path response during both training and test stages. 
\end{remark}

\begin{remark}
\label{remk: computation complexity}
Supported by the parallel computing devices, the path alignment operations can be executed in parallel. Thus, the practical runtime of path alignment operations approximately remains under typical channel matrix dimensions and oversampling factors. 
\end{remark}

\section{Training and Evaluation of Path-Oriented DL Extrapolator with Path Alignment}
In Sec.~\ref{subsec: training}, the model training is detailed to facilitate the path-oriented DL extrapolator and path alignment. After model training, the model evaluation in the test stage is discussed in Sec.~\ref{subsec: training}, which yields the extrapolated channel. 

\subsection{Training Dataset Preparation}
\label{subsec: dataset}
\begin{figure*}[t]
        \centering
        \includegraphics[width=0.9\textwidth]{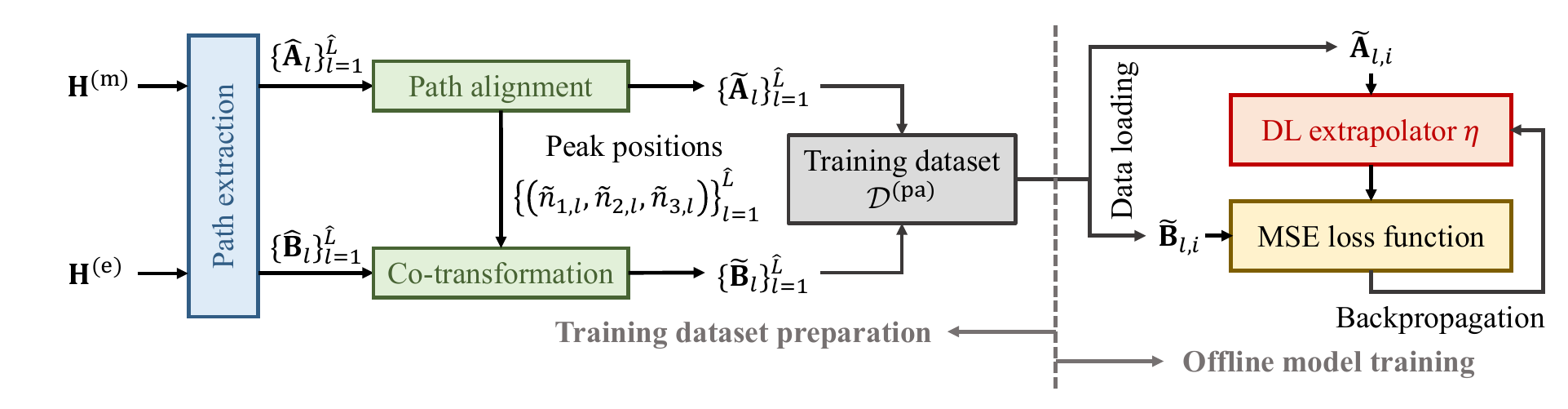}
    \captionsetup{font=footnotesize}
    \caption{End-to-end illustration of the offline training procedure of the path-oriented DL extrapolator with path alignment, which includes training dataset preparation (left) and model training (right).}
    \label{fig: training}
    \vspace{-10pt}
\end{figure*}

\begin{figure*}[t]
        \centering
        \includegraphics[width=0.8\textwidth]{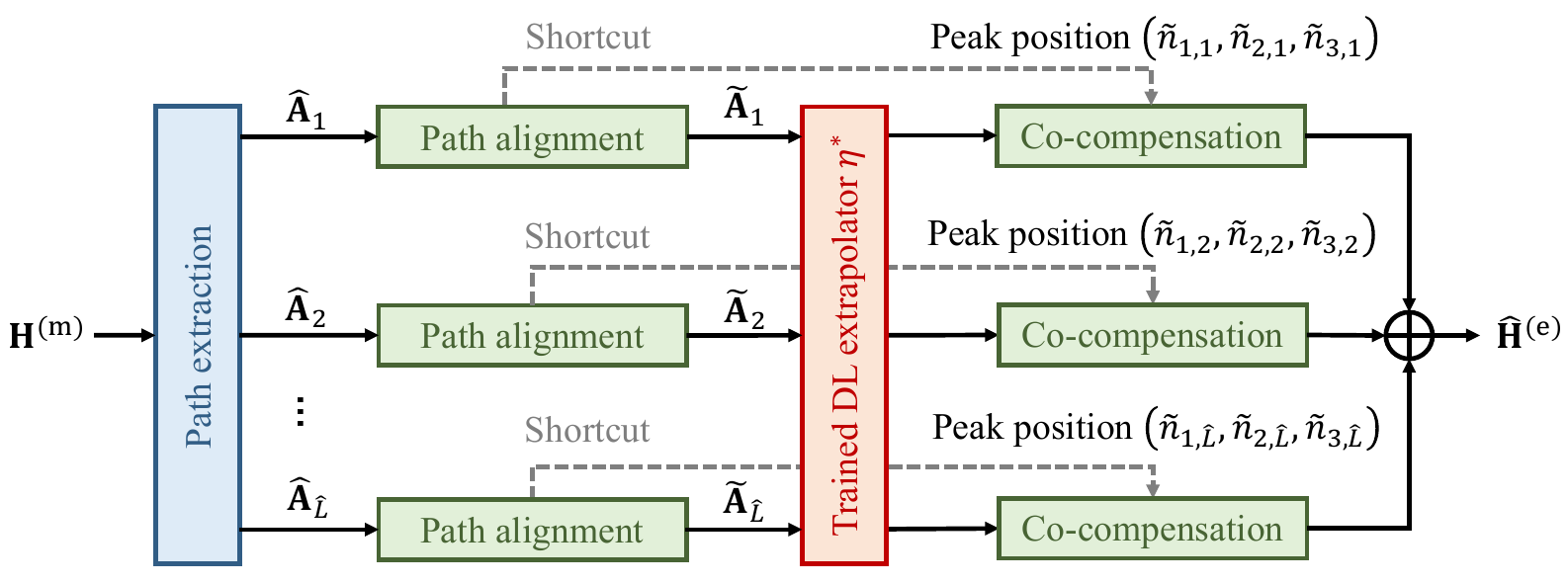}
    \captionsetup{font=footnotesize}
    \caption{Pipeline of proposed path-oriented DL extrapolator with path alignment during the online model inference stage. }
    \label{fig: structure}
    \vspace{-10pt}
\end{figure*}

The training dataset should be reformulated to facilitate the path-oriented DL extrapolator with path alignment, which is illustrated on the left of Fig.~\ref{fig: training}.  To support path-oriented design, the extracted path response $\widehat{\mathbf{A}}_{l}$ and its label $\widehat{\mathbf{B}}_{l}$ in the target frequency band are required. The generation of $\widehat{\mathbf{A}}_{l}$ has been detailed in Sec.~\ref{subsec: pe formulation}. To obtain the label $\widehat{\mathbf{B}}_{l}$, we firstly apply SAGE algorithm to $\mathbf{H}^{(\rm e)}$ to output estimated parameters $\widehat{\Theta}^{(\rm e)}=\{\hat{\alpha}_{k}^{(\rm e)},\hat{\varphi}_{k}^{(\rm e)},\hat{\theta}_{k}^{(\rm e)},\hat{\tau}_{k}^{(\rm e)}\}_{k=1}^{\hat{L}^{(\rm e)}}$. Due to the spatial reciprocity in different frequency bands, we can simultaneously cluster the estimated parameters in $\widehat{\Theta}^{(\rm e)}$ and $\widehat{\Theta}^{(\rm m)}$. Then, the label $\widehat{\mathbf{B}}_{l}$ is yielded by 
\begin{equation}
    \label{equ: label}
    \widehat{\mathbf{B}}_{l}\!=\!\sum_{k\in\mathcal{I}^{(\rm e)}_{l}}\hat{\alpha}_{k}^{(\rm e)}e^{-{\rm j}2\pi f_{1}^{(\rm e)}\hat{\tau}_{k}^{(\rm e)}}\mathbf{a}(\hat{\varphi}_{k}^{(\rm e)}, \hat{\theta}_{k}^{(\rm e)})\mathbf{b}_{\rm e}^{H}(\hat{\tau}_{k}^{(\rm e)}),
\end{equation}
where $\mathcal{I}^{(\rm e)}_{l}$ denotes the index subset of $l$th cluster. Different from the model-based channel extrapolation \cite{twc_Rottenberg_2020_performance}, the label $\widehat{\mathbf{B}}_{l}$ is directly extracted from channel $\mathbf{H}^{(\rm e)}$, whose accuracy can also be guaranteed by the monotonicity property of the SAGE algorithm and the DBSCAN clustering. Hence, the original training dataset $\mathcal{D}^{(\rm c)}$ can be reformulated into 
\begin{equation}
    \label{equ: reformulate training dataset}
    \mathcal{D}^{(\rm p)}=\left\{(\widehat{\mathbf{A}}_{l,i}, \widehat{\mathbf{B}}_{l,i})|l=1,\ldots,\hat{L}_{i},~i=1,\ldots, M\right\},
\end{equation}
where $\hat{L}_{i}$ denotes the number of extracted paths of $\mathbf{H}^{(\rm m)}_{i}$. To guarantee the learning of the downstream DL extrapolator, there are two main requirements for the extracted path responses $(\widehat{\mathbf{A}}_{l},\widehat{\mathbf{B}}_{l})$. On the one hand, the power portion of the undetected physical paths should be small. On the other hand, the extracted path response should accurately associate with the response of an individual physical path or multiple physical paths. Thus, two novel metrics are required to evaluate the performance of path extraction with respect to missed detection and physical-association fidelity. Firstly, normalized missed-detection error (NMDE) is defined as 
\begin{equation}
    \label{equ: NMDE}
    \text{NMDE}\!=\!\frac{1}{2}\!\left(\frac{\Vert\mathbf{H}^{(\rm m)}\!-\!\sum_{l=1}^{\hat{L}}\widehat{\mathbf{A}}_{l}\Vert_{F}^2}{\Vert\mathbf{H}^{(\rm m)}\Vert_{F}^2}\!+\!\frac{\Vert\mathbf{H}^{(\rm e)}\!-\!\sum_{l=1}^{\hat{L}}\widehat{\mathbf{B}}_{l}\Vert_{F}^2}{\Vert\mathbf{H}^{(\rm e)}\Vert_{F}^2}\right).
\end{equation}
As NMDE approaches zero, the impact of undetected physical paths becomes increasingly negligible. Secondly, the upper bound of normalized physical-association error (UB-NPAE) is defined as
\begin{equation}
    \label{equ: UB-NPAE-1}
    \begin{aligned}
    \text{UB-NPAE}=\frac{1}{2}\Bigg(\frac{\sum_{l=1}^{\hat{L}}\Vert\widehat{\mathbf{A}}_{l}-\sum_{k\in\mathcal{K}_{l}^*}\mathbf{A}_{k}\Vert_{F}^2}{\Vert\mathbf{H}^{(\rm m)}\Vert_{F}^2}+\\\frac{\sum_{l=1}^{\hat{L}}\Vert\widehat{\mathbf{B}}_{l}-\sum_{k\in\mathcal{K}_{l}^*}\mathbf{B}_{k}\Vert_{F}^2}{\Vert\mathbf{H}^{(\rm e)}\Vert_{F}^2}\Bigg),
    \end{aligned}
\end{equation}
where $\mathbf{B}_{k}=\alpha_{k}e^{-{\rm j}2\pi f_{1}^{(\rm e)}\tau_{k}}\mathbf{a}(\varphi_{k},\theta_{k})\mathbf{b}^{H}(\tau_{k})$ denotes the response of $k$th physical path at the target frequency band, $\{\mathcal{K}_{1}^{*},\mathcal{K}_{2}^{*},\ldots,\mathcal{K}_{\hat{L}}^*\}$ denotes a weak partition of the set $\{1, 2,\ldots,L\}$, i.e., $\mathcal{K}_{l_{1}}^*\cap\mathcal{K}_{l_{2}}^*=\emptyset,~\bigcup_{l=1}^{\hat{L}}\mathcal{K}_{l}^*=\{1,2,\ldots,L\}$ and $\mathcal{K}_{l}^*$ can be empty. The derivations of UB-NPAE are detailed in Appendix~I. When UB-NPAE approaches zero, it indicates that the extracted response $(\widehat{\mathbf{A}}_{l},\widehat{\mathbf{B}}_{l})$ can accurately reflect the physical multipath components. 

Additionally, the labels in the training dataset $\mathcal{D}^{(\rm p)}$ should be further co-transformed to facilitate the path alignment. To preserve the underlying functional relationship between input and output, the co-transformed path response $\widetilde{\mathbf{B}}_{l}$ at the target frequency band should keep the same mapping relationship in \eqref{equ: channel extrapolation} to the aligned response $\widetilde{\mathbf{A}}_{l}$. Based on the knowledge of the wideband massive MIMO channel model, the co-transformed path response $\widetilde{\mathbf{B}}_{l}$ is given as follows. 
\begin{proposition}
    \label{prop: transformed label}
    With aligned path response $\widetilde{\mathbf{A}}_{l}$ of the measured frequency band in \eqref{equ: aligned feature}, the co-transformed label $\widetilde{\mathbf{B}}_{l}$ for training can be represented by
    \begin{equation}
        \label{equ: transformed label}
        \begin{aligned}
        \widetilde{\mathbf{B}}_{l}&=\underbrace{e^{{\rm j}\beta(\tilde{n}_{3,l})}}_{\text{phase rotation}}\underbrace{\big(\mathbf{S}_{\rm a}^{(\rm e)}(\tilde{n}_{1,l},\tilde{n}_{2,l})\odot\mathbf{S}_{\rm d}^{(\rm e)}(\tilde{n}_{3,l})\big)}_{\text{angular-delay domain peak position alignment}}\odot\widehat{\mathbf{B}}_{l},\\
        &\triangleq\mathbf{V}_{l}\odot\widehat{\mathbf{B}}_{l},
        \end{aligned}
    \end{equation}
    where $\mathbf{S}_{\rm a}^{(\rm e)}(\tilde{n}_{1,l},\tilde{n}_{2,l})=\text{conj}\big(\mathbf{w}^{(\rm h,v)}_{\tilde{n}_{1,l},\tilde{n}_{2,l}}\big)\otimes\mathbf{1}_{K_{\rm e}}^{T}$, $\mathbf{S}_{\rm d}^{(\rm e)}(\tilde{n}_{3,l})=\mathbf{1}_{N_{\rm T}}\otimes(\mathbf{w}_{\tilde{n}_{3,l}}^{(\rm e)})^{T}$,
    \begin{equation}
        \label{equ: delay response target}
        \mathbf{w}_{\tilde{n}_{3,l}}^{(\rm e)}=\Big[1,e^{{\rm j}2\pi\frac{\tilde{n}_{3,l}}{O_{\rm d}K_{\rm m}}},\ldots,e^{{\rm j}2\pi\frac{\tilde{n}_{3,l}(K_{\rm e}-1)}{O_{\rm d}K_{\rm m}}}\Big]^{T},
    \end{equation}
    and 
    \begin{equation}
        \label{equ: beta}
        \beta(\tilde{n}_{3,l})=2\pi(f_{1}^{(\rm e)}-f_{1}^{(\rm m)})\frac{\tilde{n}_{3,l}}{O_{\rm d}K_{\rm m}\Delta f}. 
    \end{equation}
\end{proposition}
\begin{IEEEproof}
    See Appendix~\ref{appdix: transformation proof}.
\end{IEEEproof}
Based on \eqref{equ: transformed label}, the co-transformation of labels includes two parts. The first one is the angular-delay domain peak position alignment, which is derived from the spatial reciprocity for massive MIMO channels in different frequency bands. The second one is the phase rotation term, which is derived based on the frequency-domain channel dependency with path delays and is completely determined by the peak position $\tilde{n}_{3,l}$ in the delay domain. Then, the training dataset for path-oriented DL extrapolator with path alignment should be further reformulated from $\mathcal{D}^{(\rm p)}$ to $\mathcal{D}^{(\rm pa)}$ as 
\begin{equation}
    \label{equ: aligned dataset}
    \mathcal{D}^{(\rm pa)}=\left\{(\widetilde{\mathbf{A}}_{l,i}, \widetilde{\mathbf{B}}_{l,i})|l=1,\ldots,\hat{L}_{i},~i=1,\ldots, M\right\},
\end{equation}
where $(\widetilde{\mathbf{A}}_{l,i}, \widetilde{\mathbf{B}}_{l,i})$ denotes the 
aligned response of $(\widehat{\mathbf{A}}_{l,i},\widehat{\mathbf{B}}_{l,i})$ in $\mathcal{D}^{(\rm p)}$ based on \eqref{equ: aligned feature} and \eqref{equ: transformed label}, respectively. 

\subsection{Offline Model Training}
\label{subsec: training}
After the preparation of training dataset $\mathcal{D}^{(\rm pa)}$, neural network $\eta$ is trained by minimizing MSE loss over $\mathcal{D}^{(\rm pa)}$, which is illustrated on the right of Fig.~\ref{fig: training}. During the training, a batch of training samples $\{\{(\widetilde{\mathbf{A}}_{l,i}, \widetilde{\mathbf{B}}_{l,i})\}_{l=1}^{\hat{L}_{i}}\}_{i\in\mathcal{V}}$ is loaded from dataset $\mathcal{D}^{(\rm pa)}$, where the MSE loss function over the batch $\mathcal{V}$ is calculated by $\frac{1}{\sum_{i\in\mathcal{V}}\hat{L}_{i}}\sum_{i\in\mathcal{V}}\sum_{l=1}^{\hat{L}_{i}}\Vert\widetilde{\mathbf{B}}_{l,i}-\eta(\widetilde{\mathbf{A}}_{l,i})\Vert_{F}^2$. Then, the neural network $\eta$ is gradually optimized via backpropagation, where the trained model $\eta^*$ is yielded after convergence. Thanks to the decomposability in \eqref{equ: decompse} and the functional preservation in \eqref{equ: transformed label}, the neural network $\eta$ can adopt the same structure as the channel-oriented DL extrapolator.

\subsection{Online Model Inference}
\label{subsec: test}
The pipeline of the proposed path-oriented DL extrapolator with path alignment during the test stage is shown in Fig.~\ref{fig: structure}. Different from the model training stage, we need to output the extrapolated channel containing multiple paths for evaluation. Firstly, the path responses $\{\widehat{\mathbf{A}}_{l}\}_{l=1}^{\hat{L}}$ are extracted from $\mathbf{H}^{(\rm m)}$ based on the path parameters estimation and clustering in Sec.~\ref{subsec: pe formulation}. Then, path alignment is parallelly applied to extracted path responses $\{\widehat{\mathbf{A}}_{l}\}_{l=1}^{\hat{L}}$, which yields aligned path response $\{\widetilde{\mathbf{A}}_{l}\}_{l=1}^{\hat{L}}$ and their peak positions $\{(\tilde{n}_{1,l},\tilde{n}_{2,l},\tilde{n}_{3,l})\}_{l=1}^{\hat{L}}$. Then, the aligned path responses are parallelly input into the trained DL extrapolator $\eta^{*}$. To obtain the extrapolated channel, we should co-compensate the output of the DL extrapolator with the peak positions $\{(\tilde{n}_{1,l},\tilde{n}_{2,l},\tilde{n}_{3,l})\}_{l=1}^{\hat{L}}$ from the path alignment, which is plotted as shortcuts in Fig.~\ref{fig: structure}. Explicitly, output $\eta^{*}(\widetilde{\mathbf{A}}_{l})$ should apply phase adjustment inversely to $\mathbf{V}_{l}$, which can relocate its peak position in the angular-delay domain and compensate the phase rotation term. Then, the extrapolated channel $\widehat{\mathbf{H}}^{(\rm e)}$ is yielded by summing the co-compensated path responses as 
\begin{equation}
    \label{equ: compensation}
    \widehat{\mathbf{H}}^{(\rm e)}=\sum_{l=1}^{\hat{L}}\text{conj}(\mathbf{V}_{l})\odot\eta^{*}(\widetilde{\mathbf{A}}_{l}).
\end{equation}

\section{Theoretical Analysis of Path-Oriented DL Extrapolator with Path Alignment}
\label{subsec: advantage}
In this section, we aim to theoretically justify the contribution of distribution alignment to the environmental generalizability for path-oriented DL extrapolators with path alignment. After deployment in the test environment, the extrapolation error of the path-oriented DL extrapolator with path alignment can be calculated by 
\begin{equation}
    \label{equ: J}
    \mathcal{J}(\eta^{*})\!=\!\mathbb{E}_{Q^{(\rm pc)}}\{\Vert\widehat{\mathbf{H}}^{(\rm e)}\!-\!\mathbf{H}^{(\rm e)}\Vert_{F}^{2}\},
\end{equation}
where $(\{\widehat{\mathbf{A}}_{l}\}_{l=1}^{\hat{L}},\mathbf{H}^{(\rm e)})\sim\ Q^{(\rm pc)}$ denotes the the joint distribution in the test environment. Then, we aim to analyze the upper bound of $\mathcal{J}(\eta^{*})$. Denote $P_{\widetilde{\mathbf{A}}}^{(\rm pa)}$ and $Q_{\widetilde{\mathbf{A}}}^{(\rm pa)}$ as the marginal distribution of extracted path response at the measured frequency band after path alignment in the training and test environment. Firstly, three assumptions are made below.
\begin{assumption}
\label{asup: boundded support}
 Supports of distribution $P_{\widetilde{\mathbf{A}}}^{(\rm pa)}$ and $Q_{\widetilde{\mathbf{A}}}^{(\rm pa)}$ are both bounded by a radius $R_{1}>0$. 
\end{assumption}

\begin{assumption}
\label{asup: Lip}
Target mapping $\psi$ and DL extrapolators in the hypothesis space $\mathcal{F}$ are $R_{2}$-Lipschitz continuous. 
\end{assumption}
\begin{assumption}
\label{asup: unconditional}
The conditional expectation $\mathbb{E}\{\Vert\widetilde{\mathbf{B}}_{l}-\eta^{*}(\widetilde{\mathbf{A}}_{l})\Vert_{F}^{2}|\hat{L}\}$ is insensitive to $\hat{L}$ and can be approximated by $\mathbb{E}\{\Vert\widetilde{\mathbf{B}}_{l}-\eta^{*}(\widetilde{\mathbf{A}}_{l})\Vert_{F}^{2}|\hat{L}\}\approx\mathbb{E}\{\Vert\widetilde{\mathbf{B}}_{l}-\eta^{*}(\widetilde{\mathbf{A}}_{l})\Vert_{F}^{2}\}$.
\end{assumption}
Secondly, the Wasserstein-1 distance between two distributions is defined as follows. 
\begin{definition}
Wasserstein-1 distance $W_{1}(\mu,\nu)$ of two distributions $\mu$ and $\nu$ is defined as \cite[Definition 6.1]{villani2009optimal}
\begin{equation}
    \label{equ: wasserstein-1 def}
    W_{1}(\mu,\nu)\triangleq\inf\limits_{\gamma\sim\Gamma(\mu,\nu)}\left(\mathbb{E}_{(\widetilde{\mathbf{A}}_{1},\widetilde{\mathbf{A}}_{2})\sim\gamma}\{\Vert\widetilde{\mathbf{A}}_{1}-\widetilde{\mathbf{A}}_{2}\Vert_{F}\}\right),
\end{equation}
where $\Gamma(\mu,\nu)$ denotes all possible couplings of distribution $\mu,\nu$. $\gamma$ denotes the joint distribution of $(\widetilde{\mathbf{A}}_{1},\widetilde{\mathbf{A}}_{2})$, whose marginals are $\mu$ and $\nu$, respectively. 
\end{definition}
Then, the upper bound of $\mathcal{J}(\eta^{*})$ is given as follows. 
\begin{theorem}
\label{theo: bound original}
    Assume that $\mathcal{D}^{(\rm pa)}$ is yielded from distribution $P^{(\rm pa)}$. Then, $\mathcal{J}(\eta^{*})$ can be upper bounded by 
    \begin{equation}
        \label{equ: upper bound JQtilde}
        \begin{aligned}
        &\mathcal{J}(\eta^{*})\leq\Big(\mathcal{L}_{\mathcal{D}^{(\rm pa)}}(\eta^{*})+\underbrace{\sup_{\eta\in\mathcal{F}}|\mathcal{L}_{\mathcal{D}^{(\rm pa)}}(\eta)\!-\!\mathcal{L}_{P^{(\rm pa)}}(\eta)|}_{\text{intra-environment generalization gap}}\\
        &+\underbrace{C W_{1}\big(P_{\widetilde{\mathbf{A}}}^{(\rm pa)},Q_{\widetilde{\mathbf{A}}}^{(\rm pa)}\big)}_{\text{inter-environment generalization gap}}\Big)\times\mathbb{E}_{Q_{\hat{L}}}\{\hat{L}^{2}\},
        \end{aligned}
    \end{equation}
    where $Q_{\hat{L}}$ denotes the distribution of the number of extracted paths in the test environment and constant $C=8R_{1}R_{2}^2$. 
\end{theorem}
\begin{IEEEproof}
    See Appendix~\ref{appdix: bound 1}.
\end{IEEEproof}

\begin{remark}
\label{remk: dominance}
Under the proper path extraction algorithm and the label compensation, the underlying mapping $\widetilde{\mathbf{B}}_{l}=\psi(\widetilde{\mathbf{A}}_{l})$ can still be assumed. Thus, minimal training loss $\mathcal{L}_{\mathcal{D}^{(\rm pa)}}(\eta^{*})$ can be small enough with the powerful fitting capability of DNN and adequate optimization in training. Additionally, the intra-environment generalization gap can be reduced to 0 asymptotically by increasing dataset size \cite[Theorem 2]{twc_hu_2021_Deep}. Compared to the aforementioned two components, the inter-environment generalization gap is an intrinsic obstacle that limits environmental generalizability. 
\end{remark}
\begin{remark}
\label{remk: advantage}
For the proposed path-oriented DL extrapolator with path alignment, the inter-environment generalization gap is controlled by the distribution distance $W_{1}(P_{\widehat{\mathbf{A}}}^{(\rm pa)}, Q_{\widehat{\mathbf{A}}}^{(\rm pa)})$. Thanks to the progressive distribution alignment by path-oriented design and path alignment, a much smaller inter-environment generalization gap can be achieved compared to the channel-oriented DL extrapolator, which guarantees extrapolation performance in the unseen test environment.  
\end{remark}
\begin{remark}
    \label{remk: structure}
    In the design and the theoretical analysis, we do not constrain the specific DNN structure of DL extrapolators. Therefore, the environmental generalizability of path-oriented DL extrapolator with path alignment can be achieved by various DNN structures, e.g., MLP \cite{tcom_yang_2020_transfer} and LSTM \cite{openj_jiang_2020_deep}. 
\end{remark}

\section{Simulation Study and Sim-to-Real Discussion}
We first present the simulation setup in Sec.~\ref{subsec: setup}. Next, extensive simulation results are provided in Sec.~\ref{subsec: sim results} to validate the effectiveness of the path-oriented DL extrapolator with path alignment. Then, possible solutions to bridge `sim-to-real' gap is discussed in Sec.~\ref{subsec: sim2real}.
\subsection{Simulation Setup}
\begin{figure}[t]
        \centering
        \includegraphics[width=0.45\textwidth]{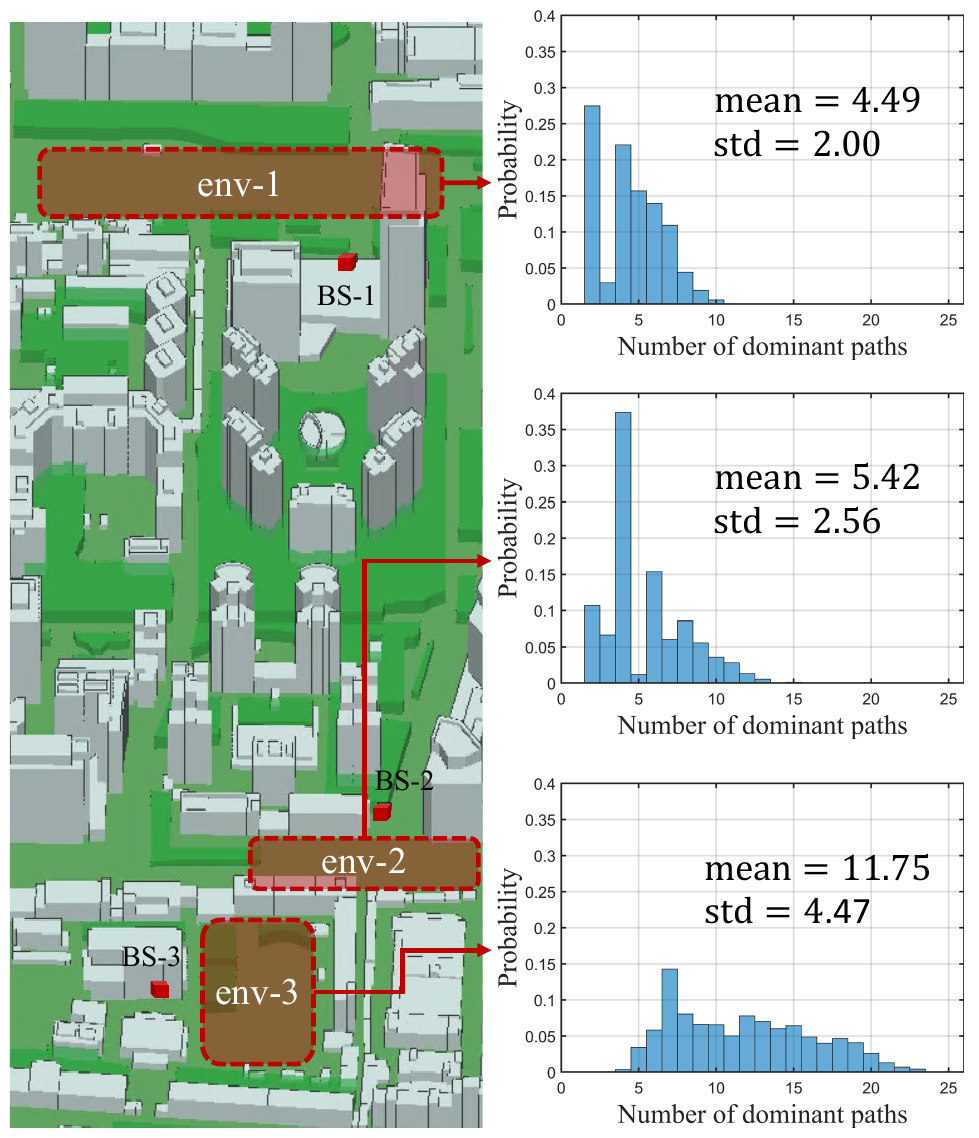}
    \captionsetup{font=footnotesize}
    \caption{Adopted urban scenario in Wireless Insite is shown on the left side, where users in env-$i$ are distributed in the transparent rectangular blocks served by BS-$i$ plotted as a red cube ($i=1,2,3$). Histograms of the number of dominant paths are shown on the right side, which is defined as the minimum number of paths that account for 99.9\% of the channel power.}
    \label{fig: sim_env}
    \vspace{-10pt}
\end{figure}
\label{subsec: setup}
In the simulation, we adopt the precise ray-tracing tool Wireless Insite \cite{wi} to generate CSI samples in different environments. The considered urban scenario is shown on the left of Fig.~\ref{fig: sim_env}, which is built on a real-world city and contains three distinct environments, namely, env-1, env-2, and env-3. It can be found that object densities vary among the three environments, where env-1 achieves the lowest object density while env-3 exhibits the highest object density. As illustrated in the right of Fig.~\ref{fig: sim_env}, the distribution of the number of dominant paths varies among the three environments. Meanwhile, due to distinct object layouts and user positions, multipath dependency and the marginal distribution of path response shift as well, which verifies \textbf{Remark}~\ref{remk: distribution shift}. 

The system configurations are detailed as follows. Starting frequencies of the measured and target frequency bands are set as $f_{1}^{(\rm m)}=3.4$ GHz and  $f_{1}^{(\rm e)}=3.5$ GHz, respectively. Numbers of subcarriers in the measured and target frequency bands are set as $K_{\rm m}=K_{\rm e}=32$. Bandwidths of measured and target frequency bands are both set as 80 MHz. The number of BS antennas is set as $N_{\rm T}=128$, where the numbers of antennas in the horizontal and vertical directions are set as 16 and 8, respectively. Heights of BS-1/2/3 are set as 25/15/20 m, respectively. The heights of the users are set as 2 m. In the simulations, ground reflection is one of the prevalent causes of the unresolvable physical paths. Based on the geometrical relationship, the maximum delay difference between the ground-reflected path to its corresponding LOS path or building-reflected path is $\frac{2\times2~\text{m}}{3\times10^8~\text{m/s}}=13.3$ ns, which is close to the system resolution $\frac{1}{80~\text{MHz}}=12.5$ ns. Considering the practical horizontal distance between the user and BS, the unresolvable ground-reflected paths are prevalent.

The default learning parameters are set as follows. Identical oversampling factors $O_{\rm h}\!=\!O_{\rm v}\!=\!O_{\rm d}\!=\!O\!=\!2$ are set in different domains. In each environment, the training dataset contains 10000 channel samples and the test dataset includes 2000 channel samples, which are uniformly collected. Considering the large dynamic range of channel power within an environment, the channel samples in the dataset are normalized by $\sqrt{N_{\rm T}K_{x}}\mathbf{H}^{(x)}/\Vert\mathbf{H}^{(x)}\Vert_{F}$ for $x\in\{{\rm m}, {\rm e}\}$. During the training, Adam optimizer \cite{iclr_Kingma_2015_Adam} with 200 training epochs is adopted, where the batch size and initial learning rate are set as 64 and $10^{-4}$, respectively. For both channel-oriented DL extrapolator (CO-DLE) and proposed path-oriented DL extrapolators with path alignment (PO-DLE+PA), a 5-layer MLP with hidden dimension 4096 is adopted as the DNN structure. To evaluate the performance of channel extrapolation, we adopt normalized mean square error (NMSE) as a metric, which is defined as 
\begin{equation}
    \label{equ: nmse}
    \text{NMSE}=\mathbb{E}\left\{\frac{\Vert\widehat{\mathbf{H}}^{(\rm e)}-\mathbf{H}^{(\rm e)}\Vert_{F}^{2}}{\Vert\mathbf{H}^{(\rm e)}\Vert_{F}^{2}}\right\},
\end{equation}
During the training stage, 10\% samples in the training dataset are split for validation. Then, the DNN with minimal validation NMSE during the training stage is adopted for testing.

\subsection{Simulation Results}
\label{subsec: sim results}
\begin{figure}[t]
    \centering
    \captionsetup{font=footnotesize,justification=centering}
    \subfloat[CDF of NMDE]{\includegraphics[width=0.24\textwidth]{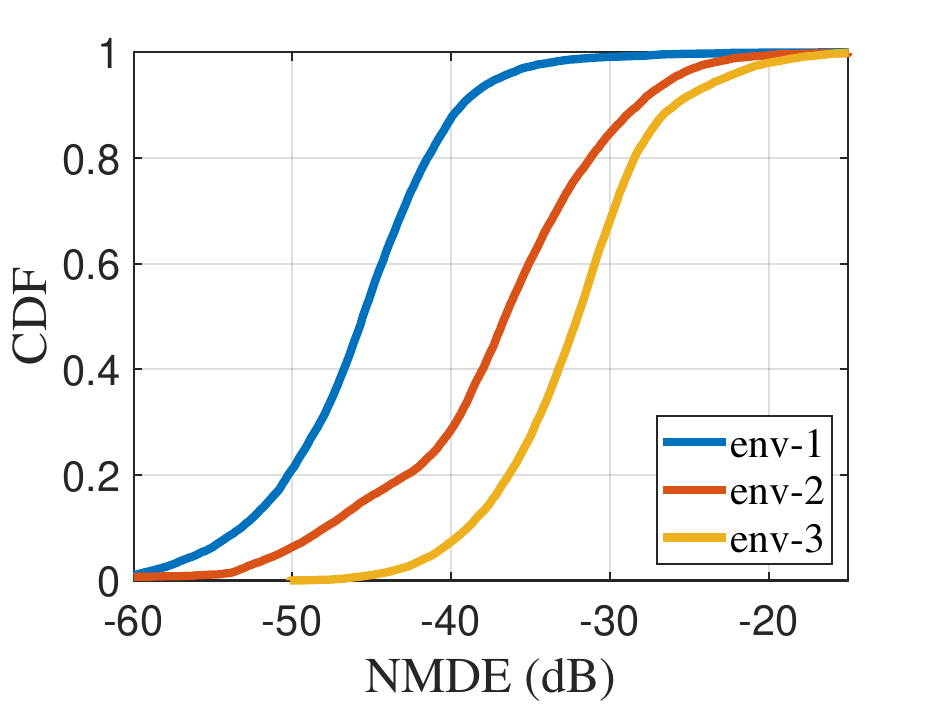}\label{subfig: NMDE}}
    \captionsetup{font=footnotesize,justification=centering}
    \subfloat[CDF of UB-NPAE]{\includegraphics[width=0.24\textwidth]{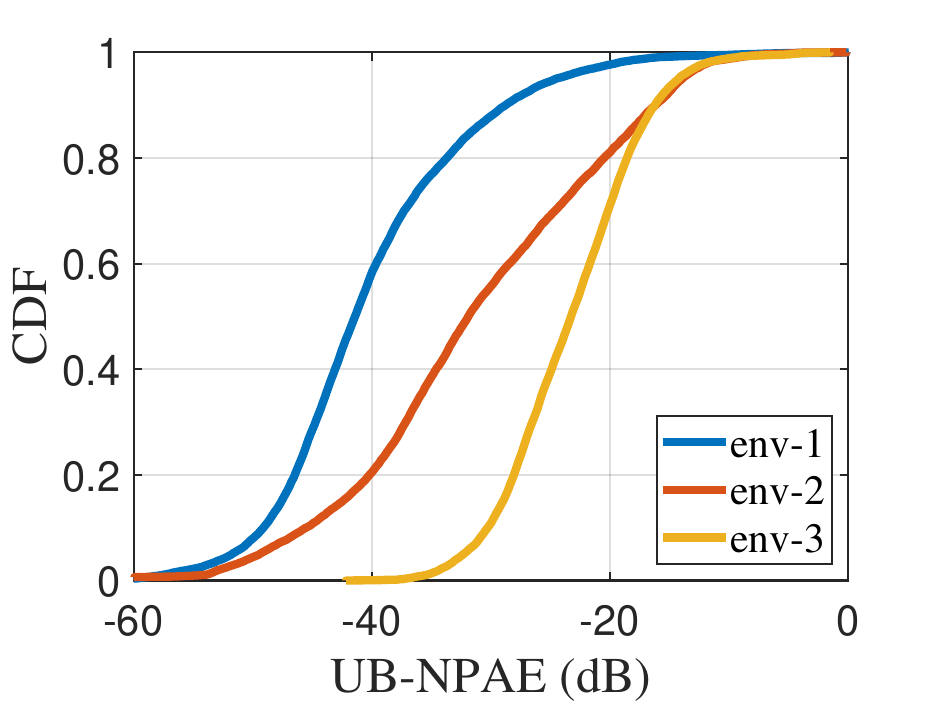}\label{subfig: UB-NPAE}}
    \captionsetup{font=small}
    \caption{Quantitative assessment of path extraction.}
    \label{fig: path extraction assessment}
    \vspace{-15pt}
\end{figure}

\subsubsection{Assessment of Path Extraction} Primarily, the NMDE and UB-NPAE of the extracted path responses in the training dataset are illustrated in Fig.~\ref{fig: path extraction assessment}. For the NMDE in env-1, env-2, and env-3, the averages are -40.9, -31.6, and -28.7 dB, respectively, with corresponding 95th percentiles of -37.0, -26.1, and -22.8 dB. The results confirm that the undetected physical paths contribute only a minor portion of the total channel power. Regarding the UB-NPAE in env-1, env-2, and env-3, the averages are -25.8, -20.6, and -19.3 dB, with 95th percentiles of -24.1, -14.1, and -14.1 dB, respectively, validating the accuracy of the physical path association. Thus, the extracted paths based on the SAGE algorithm and DBSCAN clustering can reflect real multipath components and do not significantly degrade the downstream DL extrapolation.

\subsubsection{Intra-environment Performance Study}
\label{subsubsec: val path-oriented}

\begin{figure*}[t]
    \centering
    \captionsetup{font=small}
    \subfloat[Training convergence of CO-DLE and PO-DLE+PA.]{\includegraphics[width=0.32\textwidth]{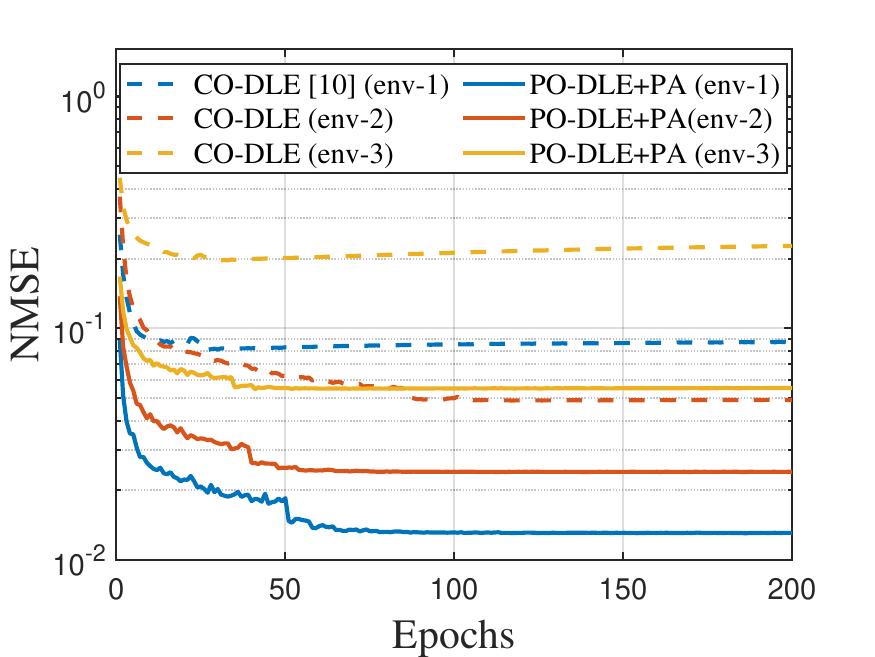}\label{subfig: train loss MLP}}
    \captionsetup{font=small}
    \subfloat[Intra-environment extrapolation performance comparison of different channel extrapolators.]{\includegraphics[width=0.32\textwidth]{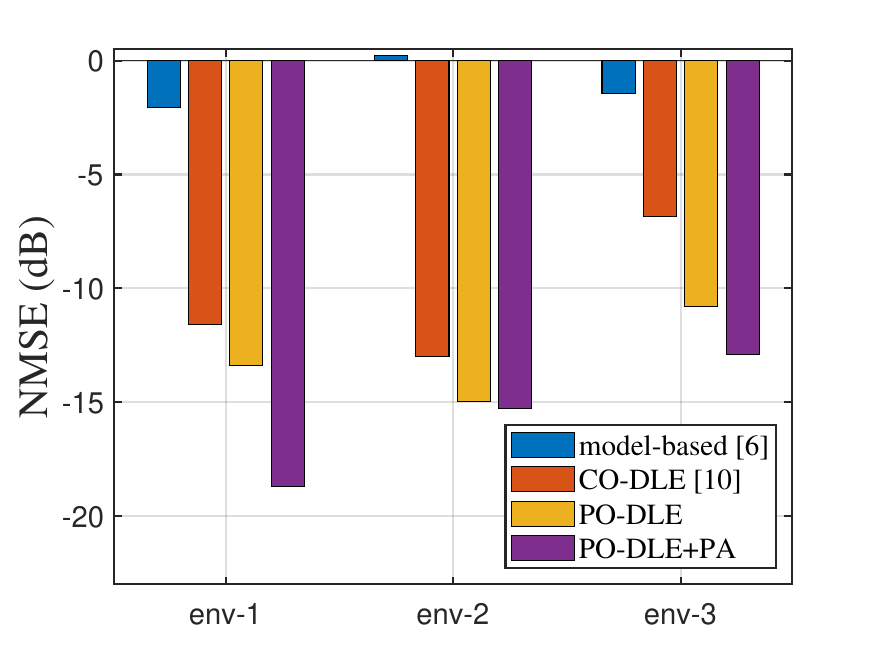}\label{subfig: intracell NMSE}}
    \captionsetup{font=small}
    \subfloat[NMSE comparison of DL extrapolators trained in env-1 under different training dataset size.]{\includegraphics[width=0.32\textwidth]{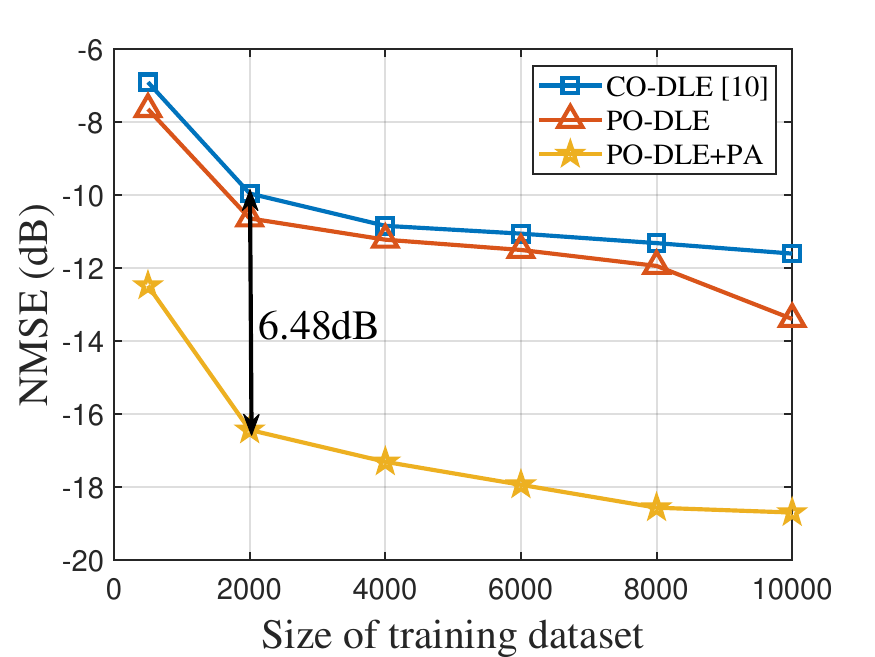}\label{subfig: intra-env versus samples env-1}}
    \captionsetup{font=small}
    \caption{Intra-environment performance verification of path-oriented DL extrapolator with path alignment.}
    \label{fig: intra-env}
    \vspace{-10pt}
\end{figure*}

\begin{figure*}[t]
    \centering
    \captionsetup{font=small,justification=centering}
    \subfloat[
    Tranined in env-3 and test in env-1/2]{\includegraphics[width=0.32\textwidth]{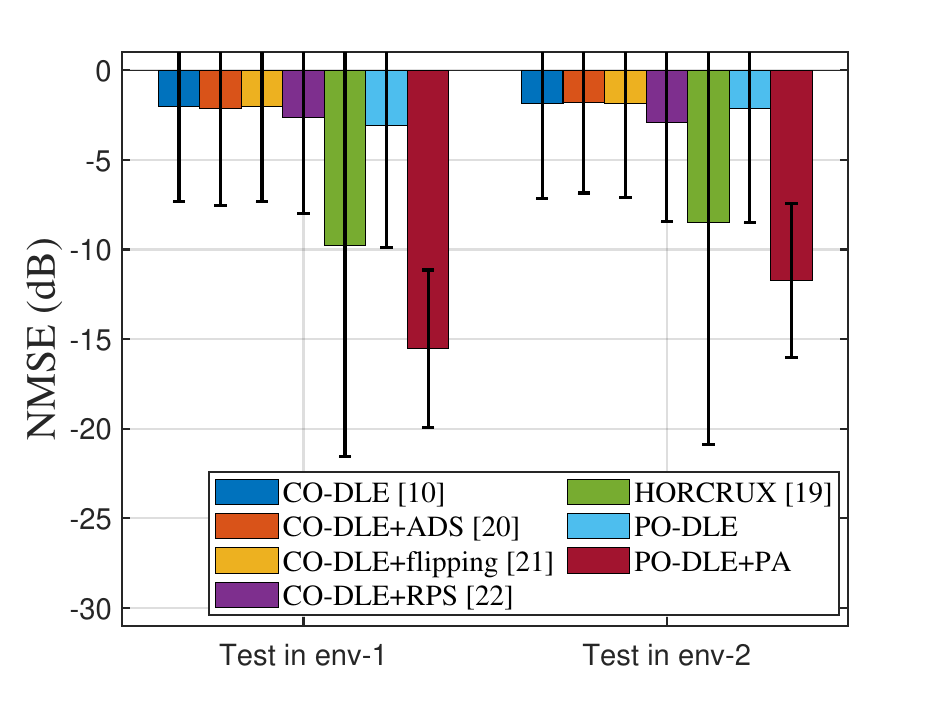}\label{subfig: inter update}}
    \captionsetup{font=small,justification=centering}
    \subfloat[Trained in env-1 and test in env-3]{\includegraphics[width=0.32\textwidth]{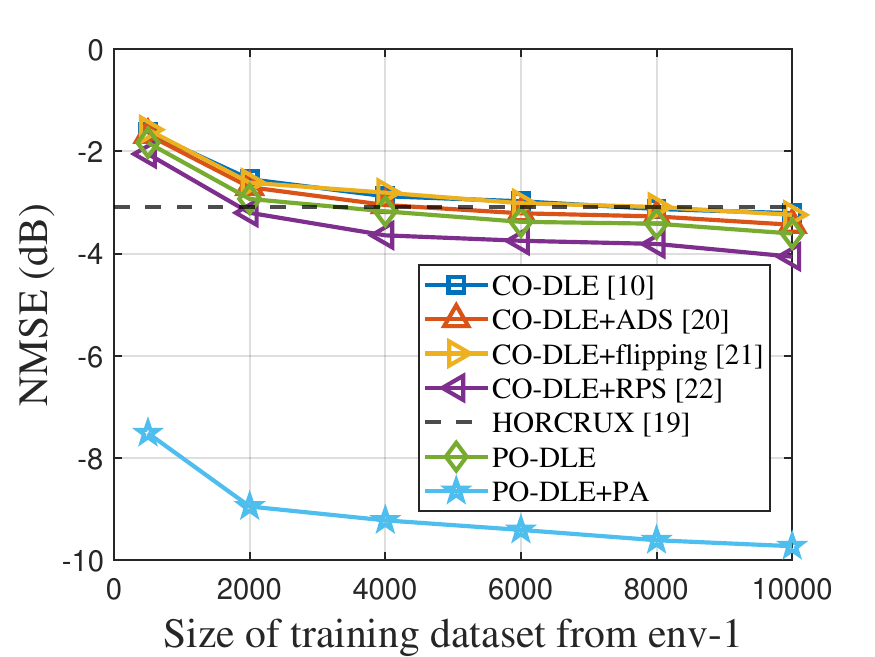}\label{subfig: pretrained env-1}}
    \captionsetup{font=small,justification=centering}
    \subfloat[Trained in env-2 and test in env-3]{\includegraphics[width=0.32\textwidth]{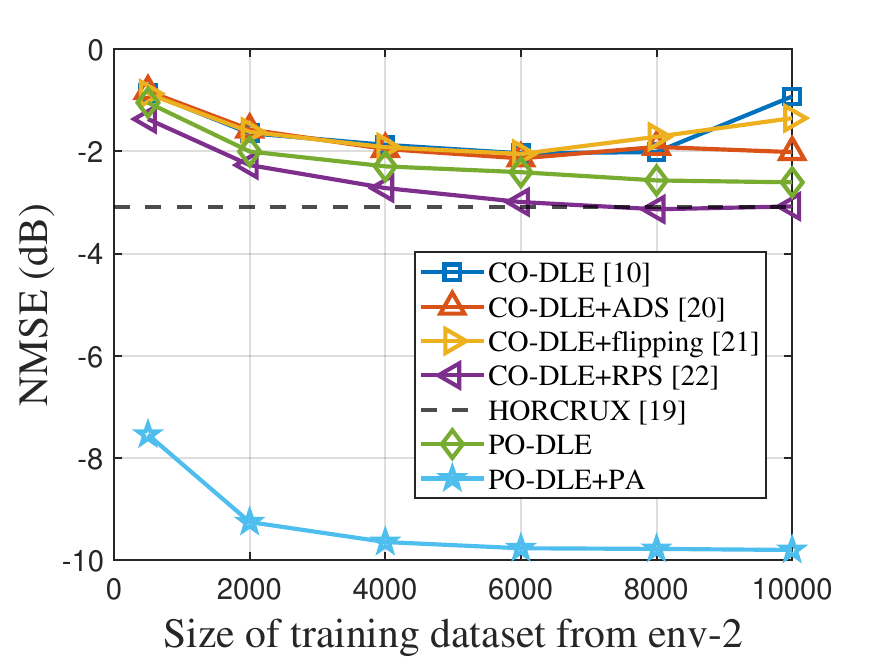}\label{subfig: pretrained env-2}}
    \captionsetup{font=small,justification=raggedright }
    \caption{Inter-environment generalizability justification of path-oriented DL extrapolator with path alignment.}
    \label{fig: inter performance}
    \vspace{-10pt}
\end{figure*}

Before the enhancement of environmental generalizability, the proposed PO-DLE+PA should guarantee intra-environment extrapolation performance. Hereby, we consider the CO-DLE as a baseline. Then, the NMSE of validation samples during the training stage is depicted in Fig.~\ref{fig: intra-env}\subref{subfig: train loss MLP}. It can be found that the validation NMSE of PO-DLE+PA can converge at a much lower NMSE compared to the CO-DLE.

Then, the trained extrapolators are tested in the same environment to verify the intra-environment performance. In addition, we adopt the model-based extrapolator \cite{twc_Rottenberg_2020_performance} as another baseline, which directly calculates the target CSI based on the estimated parameters from the SAGE algorithm in the path extraction step. Intra-environment extrapolation NMSE over three environments is illustrated in Fig.~\ref{fig: intra-env}\subref{subfig: intracell NMSE}. It can be found that the performance of the PO-DLE obviously surpasses the model-based approach, which demonstrates the necessity of DL models for accurate channel extrapolation. Meanwhile, the proposed PO-DLE can further achieve a 2.6 dB NMSE reduction on average compared to its CO-DLE counterpart. Moreover, with the path alignment approach, the intra-environment extrapolation NMSE of PO-DLE+PA can be further reduced by 2.6 dB on average. 

Subsequently, we vary the training dataset size in env-1 from 500 to 10000 and depict the test NMSE of different DL extrapolators in Fig.~\ref{fig: intra-env}\subref{subfig: intra-env versus samples env-1}. It can be found that the test NMSE of all DL extrapolators decreases with the training dataset size. The rationale lies that a smaller intra-environment generalization gap in \eqref{equ: upper bound JQtilde} can be achieved by a larger training dataset size. Additionally, an evident intra-environment performance gain of PO-DLE+PA can be achieved with path alignment when the training dataset is small. For instance, when the training dataset size equals 2000, a 6.48 dB extrapolation performance gain can be achieved by the PO-DLE+PA compared to the CO-DLE. As a result, a much smaller training dataset is required in PO-DLE+PA to guarantee intra-environment performance.

\subsubsection{Environmental Generalizability Study} 
 
In the beginning, the environmental generalizability of PO-DLE+PA is validated as follows. We consider HORCRUX extrapolator \cite{mobicom_Banerjee_2024_HORCRUX} and CO-DLE with different data augmentation schemes as the baselines. For the HORCRUX extrapolator, eight parallel neural network channel dividers and mini-neural network distance estimators are adopted, where all the neural networks are trained with the same settings in \cite{mobicom_Banerjee_2024_HORCRUX}. Explicitly, each pair of the neural network channel divider and mini-neural network distance estimator is trained with a grounded multipath channel dataset with 100000 samples, where the path parameters are randomly generated. Data augmentation schemes include random angular-delay random circular shift (ADS) \cite{twc_liu_2024_deep}, flipping \cite{chinacom_han_2024_AI} and random phase shift (RPS) \cite{jstsp_guo_2022_user}. Since these data augmentation baselines were originally designed for channel feedback, where the input and labels are the same, they are modified below to preserve the dependency between the augmented input and label for channel extrapolation. 
\begin{itemize}
    \item ADS: Shifts $(\tilde{n}_{1,l},\tilde{n}_{2,l},\tilde{n}_{3,l})$ are uniformly chosen from $[0, O_{\rm h}N_{\rm h}-1]\times[0, O_{\rm v}N_{\rm v}-1]\times[0, O_{\rm d}N_{\rm d}-1]$. Then, the augmented input and label are yielded by multiplying $\mathbf{U}_{l}$ in \eqref{equ: aligned feature} to the input and $\mathbf{V}_{l}$ in \eqref{equ: transformed label} to the label, respectively. 
    \item Flipping: Simultaneously flip the input and label matrix along its first dimension, i.e., antenna dimension. 
    \item RPS: Simultaneously multiply random phase shift term $e^{j\phi}$ to the input and label, where $\phi\sim U(0,2\pi)$. 
\end{itemize}
After offline data augmentation, the size of the training dataset is doubled compared to the original training dataset size. Firstly, the DL extrapolators are trained in env-3 and the extrapolation performances in env-1 and env-2 are shown in Fig.~\ref{fig: inter performance}\subref{subfig: inter update}. For proposed PO-DLE+PA, -15.54 dB and -11.72 dB extrapolation NMSE can be achieved in unseen env-1 and env-2, which surpasses the performance of HORCRUX by 3.3$\sim$5.8 dB. As a result, environmental generalizability can be enabled by PO-DLE+PA. In contrast, the performance of CO-DLE is degraded to -2 dB in env-2 and env-3, which is caused by the distribution shift of the wireless channel. Additionally, the extrapolation performance gain is less than 1.1 dB when data augmentation schemes are applied in CO-DLE. The rationale lies in the diversity of wireless environments, where the distribution shift of wireless channels cannot be effectively reduced via data augmentation schemes. Further, the generalization robustness of different approaches in Fig.~\ref{fig: inter performance}\subref{subfig: inter update} is also investigated. It can be observed that the standard deviations (error bars) of PO-DLE+PA are around 4.3 dB for the users randomly distributed in env-1 and env-2. Moreover, the proposed PO-DLE+PA can achieve a 95th percentile NMSE of -10.98 dB and -6.76 dB when tested in env-1 and env-2. On the contrary, the HORCRUX baseline exhibits large standard deviations of NMSE ($>10$ dB) in unseen env-1 and env-2, which is more sensitive to the user positions. Thus, the generalization robustness of PO-DLE+PA across random user placements can be justified. 

Next, we consider the environmental generalizability of DL extrapolators under different training dataset sizes from the source environment. Hereby, the DL extrapolators are trained in env-1 or env-2 and then are tested in unseen env-3. Since env-3 exhibits a richer multipath condition compared to env-1 and env-2, it is more challenging when generalizing to env-3. Then, we vary the training dataset size in env-1 and env-2 from 500 to 10000 and plot the extrapolation NMSE in Fig.~\ref{fig: inter performance}\subref{subfig: pretrained env-1} and Fig.~\ref{fig: inter performance}\subref{subfig: pretrained env-2}. For the proposed PO-DLE+PA, around -10 dB extrapolation NMSE can be realized in the unseen env-3, which has been reduced by 6 dB compared to the baselines. Thus, the proposed PO-DLE+PA can also accurately generalize to an unseen environment with a richer multipath condition. Meanwhile, since the inter-environment generalization gap has been greatly reduced in PO-DLE+PA, -9 dB generalization NMSE can be achieved with only 2000 training samples in the source environment. Thus, strong environmental generalizability with a small training dataset is also enabled in the proposed PO-DLE+PA. Comparing CO-DLE, PO-DLE, and PO-DLE+PA, their extrapolation NMSE in unseen env-3 is progressively reduced, which validates \textbf{Remark}~\ref{remk: alignment}. Additionally, the NMSE of the HORCRUX extrapolator is degraded to -3 dB due to the severe distribution shift of training and test data in env-3, which validates \textbf{Remark}~\ref{remk: knowledge function}. Meanwhile, the NMSE of the baselines almost remains and cannot fall below -4.1 dB when the training dataset size exceeds 6000. This result indicates that the inter-environment distribution gap is a fundamental limit for environmental generalizability and cannot be mitigated by increasing training dataset size, which validates \textbf{Remark}~\ref{remk: dominance}.

\subsubsection{Theoretical Consistency}
\begin{figure}[t]
    \vspace{-10pt}
        \centering
        \includegraphics[width=0.45\textwidth]{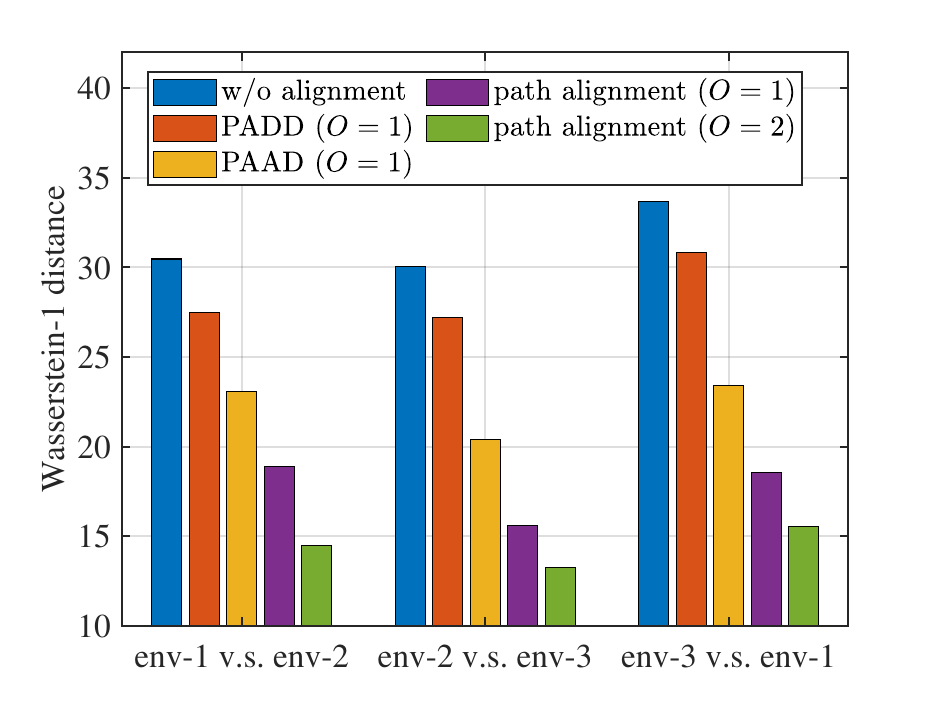}
    \vspace{-5pt}
    \captionsetup{font=footnotesize}
    \caption{Wasserstein-1 distance of extracted path response between different environments, which can quantify the distribution shift.
    }
    \label{fig: dw}
\end{figure}

\begin{table}[t]
  \centering
  \belowrulesep=0pt
  \aboverulesep=0pt
  \captionsetup{font=footnotesize}
  \caption{Extrapolation NMSE (in dB) of path-oriented DL extrapolators generalizing to unseen environments under different alignment approaches, where the best generalization performance is in bold.}
    \begin{tabular}{c|c|c|c|c|c|c}
    \toprule
    \multicolumn{1}{c|}{{trained\newline{}}} & \multicolumn{1}{c|}{{test\newline{}}} & \multicolumn{1}{c|}{{w/o\newline{}}} & \multicolumn{1}{c|}{{PADD\newline{}}} & \multicolumn{1}{c|}{{PAAD\newline{}}} & \multicolumn{1}{c|}{{PA\newline{}}} & \multicolumn{1}{c}{{PA\newline{}}} \\
    env.      &    env.   &    alignment   &    (O=1)   &   (O=1)    &   (O=1)   & (O=2) \\
    \midrule
    \multirow{2}{*}{\centering env-1} & env-2 & -4.03 & -4.85 & -5.04 & -7.24 & \textbf{-9.84} \\
\cmidrule{2-7}          & env-3 & -3.61 & -3.74 & -6.43 & -9.11 & \textbf{-9.73} \\
    \midrule
    \multirow{2}{*}{\centering env-2} & env-1 & -3.15 & -7.60 & -9.26 & -11.97 & \textbf{-13.35} \\
\cmidrule{2-7}          & env-3 & -2.61 & -3.34 & -5.74 & -9.19 & \textbf{-9.80} \\
    \midrule
    \multirow{2}{*}{\centering env-3} & env-1 & -3.1  & -3.42 & -10.40 & -13.95 & \textbf{-15.54} \\
\cmidrule{2-7}          & env-2 & -2.14 & -1.52 & -7.63 & -9.74 & \textbf{-11.72} \\
    \bottomrule
    \end{tabular}%
  \label{tab: ablation}%
  \vspace{-5pt}
\end{table}
The consistency between the inter-environment generalization gap in \textbf{Theorem}~\ref{theo: bound original} and environmental generalizability of PO-DLE+PA is investigated as follows. According to Sec.~\ref{subsec: alignment}, the path alignment approach aims to reduce the distribution shift of extracted path response, which is composed of 3 components: (1) peak position alignment in delay-domain (PADD); (2) peak position alignment in angular-domain (PAAD); (3) oversampling for power leakage compression. To quantify distribution shift reduction, we plot the Wasserstein-1 distance of the extracted path response in Fig.~\ref{fig: dw} under the combinations of the aforementioned three components. It can be found that PADD, PAAD, and their combination (path alignment, PA) can reduce the distribution shift of the extracted path response.  Meanwhile, as the oversampling factor increases from 1 to 2, the Wasserstein-1 distance can be further reduced by around 20\% by suppressing power leakage. Therefore, the path alignment approach can explicitly reduce the distribution shift of extracted path response, which validates \textbf{Remark}~\ref{remk: distinction}. Further, the environmental generalization performances of PO-DLE with different alignment approaches are listed in Table~\ref{tab: ablation}. With the decrease of Wasserstein-1 distances of path response distribution via distribution alignment, environmental generalizability of proposed PO-DLE is consistently enhanced, which proves \textbf{Remark}~\ref{remk: advantage} \footnote{With normalization in the dataset, the constant $R_{1}$ can be grounded by $R_{1}=\sqrt{N_{\rm T}K_{\rm m}}=64$. Nevertheless, the exact calculation of Lipschitz constant $R_{2}$ in DNN is NP-hard, and only a loose upper bound can be attained \cite{nips_Aladin_2018_lip}. Although the generalization upper bound in \eqref{equ: upper bound JQtilde} is loose in numerical tightness, it can still serve as a theoretical guideline to enhance generalizability by reducing Wasserstein-1 distance.}. 

\subsubsection{Flexibility and Robustness Study}
\begin{figure}[t]
    \vspace{-10pt}
        \centering
        \includegraphics[width=0.45\textwidth]{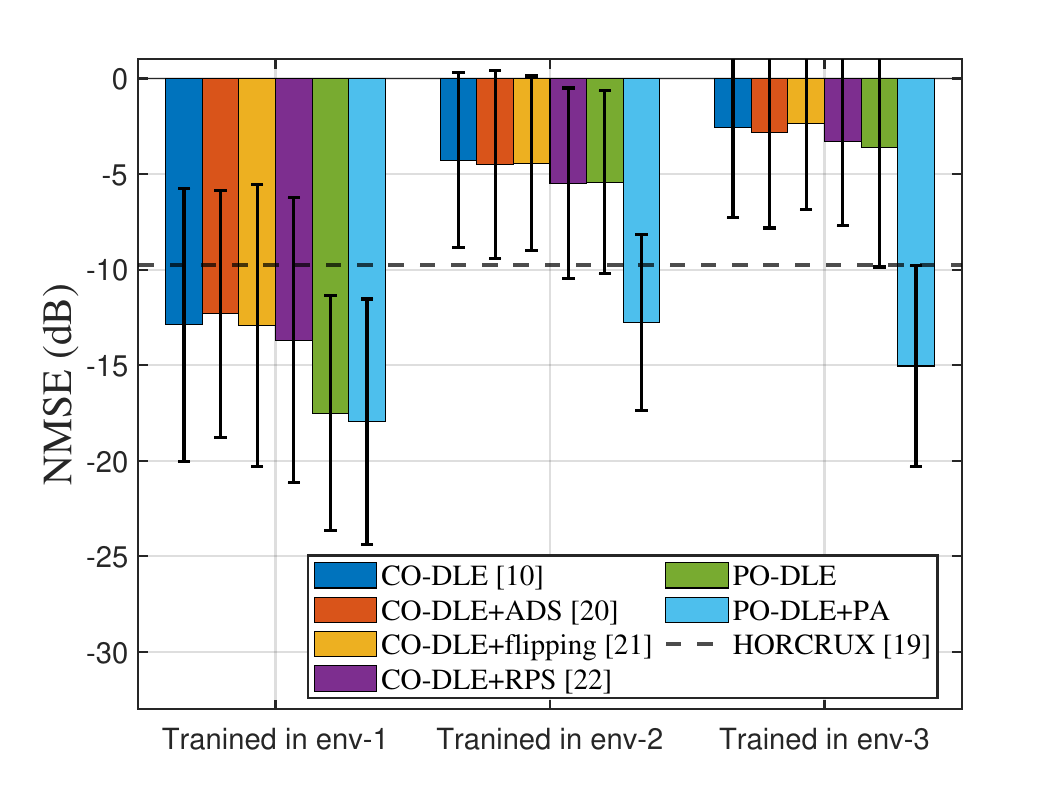}
    \vspace{-5pt}
    \captionsetup{font=footnotesize}
    \caption{Extrapolation NMSE in env-1 of DL extrapolators trained in different environments, where the sequential LSTM is adopted as the DNN structure for CO-DLE, PO-DLE, and PO-DLE+PA.}
    \label{fig: lstm}
    \vspace{-10pt}
\end{figure}
\begin{figure}[t]
        \centering
        \includegraphics[width=0.45\textwidth]{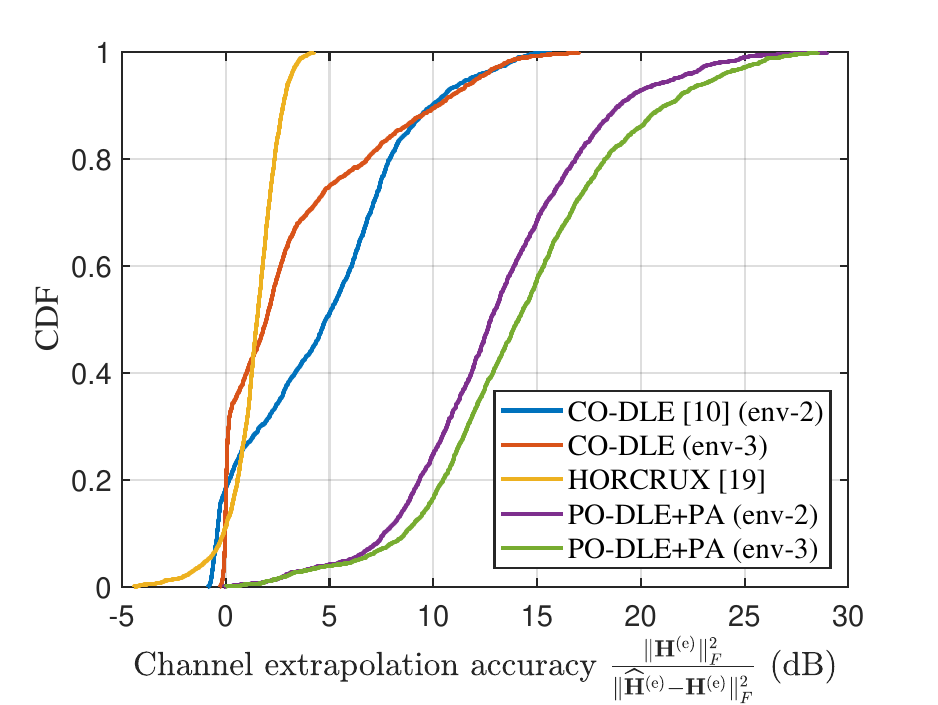}
    \vspace{-5pt}
    \captionsetup{font=footnotesize}
    \caption{CDF of channel extrapolation accuracy with $\text{SNR}=5$ dB, where DL extrapolators are tested in env-1 and training environments are referred in the legend.}
    \label{fig: SNR}
    \vspace{-10pt}
\end{figure}
The proposed PO-DLE+PA is a general and flexible framework for enhancing environment generalizability, which can be realized by various DNN structures. Thus, the flexibility of DNN structure over intra-environment performance and inter-environment generalizability is investigated below. We adopt sequential LSTM \cite{openj_jiang_2020_deep,wcl_Yao_2024_loss} as the DNN structure in the CO-DLE, PO-DLE, and PO-DLE+PA. Here, three convolutional layers are adopted for feature extraction, where the input/output channels are (2,512), (512,1024), (1024,1024), respectively. The number of hidden layers in LSTM is set as 2, where the hidden dimension is set as 1024. Then, the test NMSE in env-1 for DL extrapolators trained in different environments is depicted in Fig.~\ref{fig: lstm}. When trained in env-1, it can be found that the performance of PO-DLE surpasses the baselines by 3.7$\sim$7.74 dB. Meanwhile, intra-environment performance can still be guaranteed for PO-DLE+PA. While trained in env-2 and env-3, PO-DLE+PA can still generalize to the unseen env-1. In contrast, severe performance degradation occurs in the CO-DLE when tested in the unseen env-1. Additionally, the environmental generalizability cannot be effectively enhanced by data augmentation approaches, either. Therefore, the intra-environment performance and inter-environment generalizability of PO-DLE+PA are still held with different DNN structures, which verifies \textbf{Remark}~\ref{remk: structure}. 

Robustness against additive noise is critical in the practical operation of channel extrapolation. Thus, the environmental generalizability of DL extrapolators with noised channel samples is investigated below, where complex Gaussian noise with zero mean is added to the training and test datasets. Herby, we fix the signal-to-noise ratio (SNR) as 5 dB and consider channel extrapolation accuracy $\frac{\Vert\mathbf{H}^{(\rm e)}\Vert_{F}^{2}}{\Vert\widehat{\mathbf{H}}^{(\rm e)}-\mathbf{H}^{(\rm e)}\Vert_{F}^{2}}$ as a performance indicator under noise. Then, the cumulative distribution function (CDF) of channel extrapolation accuracy of different DL extrapolators when tested in unseen env-1 is illustrated in Fig.~\ref{fig: SNR}. When operating in a low-SNR condition with $\text{SNR}=5$ dB, the proposed PO-DLE+PA can still achieve the highest channel extrapolation accuracy, where a stable and obvious generalization gain ($>8$ dB) can be achieved in different percentiles. Compared to the noise-free condition, the extrapolation NMSE degradation of PO-DLE+PA in unseen env-1 is less than 3.4 dB on average. Thus, the robustness against noise of the proposed PO-DLE+PA can be verified. On the contrary, the extrapolation NMSE of the HORCRUX extrapolator has been degraded by more than 8.5 dB compared to the noise-free condition, which is more susceptible to additive noise. 

\begin{figure}[t]
    \vspace{-10pt}
        \centering
        \includegraphics[width=0.45\textwidth]{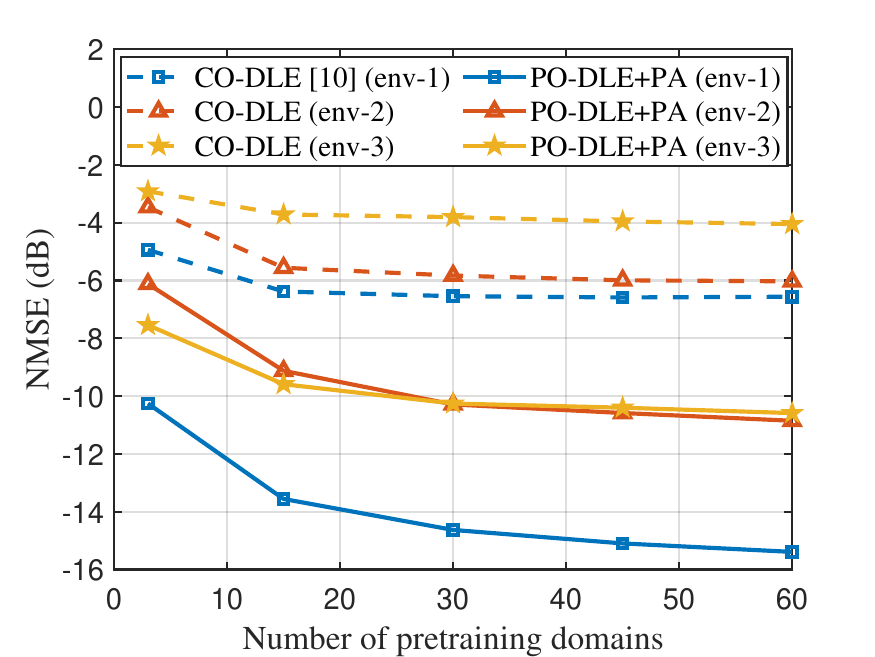}
    \vspace{-5pt}
    \captionsetup{font=footnotesize}
    \caption{Generalization comparison under different numbers of pretraining domains, where the test environments are shown in the legend.}
    \label{fig: domains}
    \vspace{-10pt}
\end{figure}
\subsubsection{Pretraining in Diverse Domains}Further, the generalizability of the DL extrapolators with diverse training datasets is investigated. To this end, we adopt the `City Scenario Series' in the DeepMIMO dataset \cite{ita_Ahmed_2019_deepmimo} to generate diverse training datasets. Explicitly, the `City Scenario Series' dataset consists of ray-tracing results in 20 real-world city scenarios, where each scenario contains three BS located at different positions. Thus, up to 60 domains can be generated for pretraining. The configurations of the BS in the pertaining dataset are aligned with Sec.~\ref{subsec: setup}. Then, the generalization NMSE of PO-DLE+PA and CO-DLE under different numbers of training domains is plotted in Fig.~\ref{fig: domains}. It can be found that the generalizability of the proposed PO-DLE+PA can be gradually enhanced by increasing the number of pretraining domains, where the extrapolation NMSE can be reduced below -15 dB in env-1 and below -10 dB in env-2 and env-3. Thus, the proposed PO-DLE+PA can also benefit from the training dataset from multiple environments to enhance generalizability. On the contrary, the generalizability of the CO-DLE nearly saturates when the number of pretraining domains exceeds 15, which cannot be effectively enhanced by further increasing training dataset diversity. Consequently, merely increasing the diversity of training domains cannot effectively address the generalizability challenge of CO-DLE in unseen environments. Based on the comparison above, PO-DLE+PA is still required for generalizability enhancement when a diverse pretraining dataset is available.
\subsubsection{Runtime Analysis} The runtime of peak position search under different oversampling factors is presented in Table~\ref{tab: peak search runtime}, which is tested on the Nvidia GeForce RTX 3090 GPU. It can be found that the peak search runtime nearly remains with the oversampling factor $O$. Therefore, the oversampling-based peak position search can enhance the environment-generalizability with affordable runtime. Further, the runtime of path alignment under different antenna sizes and number of subcarriers are considered. For both the 128-antenna system with 32 subcarriers and the 256-antenna system with 64 subcarriers, the runtime of path alignment remains 0.26 ms, which validates \textbf{Remark}~\ref{remk: computation complexity}.

\begin{table}[t]
  \centering
  \belowrulesep=0pt
  \aboverulesep=0pt
  \captionsetup{font=footnotesize}
  \caption{Runtime of oversampling-based peak search}
    \begin{tabular}{c|c|c|c|c}
    \toprule
    oversampling factor $O$ & 1     & 2     & 3     & 4 \\
    \midrule
    peak search runtime (ms) & 0.192 & 0.188 & 0.192 & 0.194 \\
    \bottomrule
    \end{tabular}%
  \label{tab: peak search runtime}%
  \vspace{-10pt}
\end{table}%

\subsection{Discussion: Bridging Sim-to-Real Gap}
\label{subsec: sim2real}
\subsubsection{Diverse pretraining dataset with high-fidelity} As presented in Fig.~\ref{fig: domains}, a diverse pretraining can enhance the generalizability of PO-DLE+PA. Additionally, realistic factors can be modeled into the ray-tracing tool and signal generation model, which can reduce the `sim-to-real' gap. Firstly, diverse scenarios can be generated to enhance the diversity of large-scale channel parameters. Secondly, a more comprehensive signal propagation mechanism with diffraction can be adopted in the ray-tracing tool. Additionally, a ray-tracing-statistical hybrid modeling \cite{twc_chen_2021_channel} can be used to generate diffusion paths. Thirdly, hardware effects, including antenna pattern and frequency response, can be modeled in the received signals \cite{sensys_chen_2023_RFGenesis}. Finally, generative models can also be adopted to generate synthetic signals with few real-world measurements \cite{mobicom_chi_2024_rf}, which can model the non-idealities in practical scenarios.
\subsubsection{Reliable path extraction in real world} Before path extraction, real-world RF system response should be compensated in the raw channel measurements \cite{twc_choi_2021_experimental}. Then, modified SAGE algorithms \cite{twc_choi_2021_experimental,twc_yin_2017_scatter} are required when pretraining in the simulated diverse dataset and deploying in the real-world scenarios, which can facilitate the path extraction performance under practical hardware effects and different propagation conditions (e.g., spherical wave propagation in near-field).

\subsubsection{Frequency-domain signal calibration}Frequency-domain signal calibration is required to facilitate the invariance of the target function in practical scenarios. Explicitly, the input and labels for the path-oriented DL extrapolator should be calibrated by dividing the antenna frequency response. Thus, the trained neural network will not fit the response of a specific antenna in the training dataset. 

\section{Conclusion} 
Under the distribution shift of wireless channels across environments, environmental generalizability is vital to reducing the large-scale deployment cost for DL extrapolators. In this paper, generalizable learning for channel extrapolation is realized. Firstly, we analyze the distribution shift of wireless channels, which comprises the distribution shift of multipath structure and single-path response. Secondly, a physics-based distribution alignment is proposed to progressively resolve the distribution shift of wireless channels, which contains the two steps of path-oriented design and path alignment. Inspired by the path-wise decomposability of channel extrapolation in physics, path-oriented DL extrapolator is proposed to parallelly extrapolate extracted paths via DL, which tackles the distribution shift of multipath structure. Then, path alignment is proposed to mitigate the distribution shift of extracted path response in path-oriented DL extrapolators. Thirdly, label co-transformation in training and output co-compensation during test are designed to facilitate path-oriented DL extrapolators with path alignment. Comprehensive simulation results reveal that the proposed path-oriented DL extrapolator with path alignment can achieve strong environmental generalizability while guaranteeing intra-environment performance. Additionally, the environmental generalizability of path-oriented DL extrapolators with path alignment is consistent with its distribution alignment metric, which exhibits strong theoretical interpretability. A more detailed and comprehensive implementation to reduce `sim-to-real' gap is left as our future work.

\appendices
\section{Derivations of UB-NPAE}
\label{appdix: UB-NPAE}
To quantify the physical-association filedlity of path extraction, the NPAE is originally defined as 
\begin{equation}
    \label{equ: NPAE-1}
    \begin{aligned}
    \text{NPAE}=\min_{\{\mathcal{K}_{1},\ldots,\mathcal{K}_{\hat{L}}\}}\Bigg\{\frac{1}{2}\Bigg(\frac{\sum_{l=1}^{\hat{L}}\Vert\widehat{\mathbf{A}}_{l}-\sum_{k\in\mathcal{K}_{l}}\mathbf{A}_{k}\Vert_{F}^2}{\Vert\mathbf{H}^{(\rm m)}\Vert_{F}^2}\\+\frac{\sum_{l=1}^{\hat{L}}\Vert\widehat{\mathbf{B}}_{l}-\sum_{k\in\mathcal{K}_{l}}\mathbf{B}_{k}\Vert_{F}^2}{\Vert\mathbf{H}^{(\rm e)}\Vert_{F}^2}\Bigg)\Bigg\},
    \end{aligned}
\end{equation}
where $\{\mathcal{K}_{1},\ldots,\mathcal{K}_{\hat{L}}\}$ denotes a weak partition of set $\{1,2,\ldots,L\}$. Based on the combinatorics, the number of all possible weak partitions is $\hat{L}^{L}$, which exponentially grows with $L$ and increases the complexity of the exact NPAE calculation. To this end, we consider UB-NPAE with low computation complexity. Explicitly, a greedy weak partition $\{\mathcal{K}_{1}^*,\ldots,\mathcal{K}_{\hat{L}}^*\}$ is heuristically determined based on the distance between the physical paths and extracted path responses. Consider the distance $D(\boldsymbol{\omega}_{k},\hat{\boldsymbol{\omega}}_{l}^{(\rm p)})$ between the $k$th physical path and $l$th extracted path response, where $\boldsymbol{\omega}_{k}=\big[\pi\sin{(\varphi_{k})}\sin{(\theta_{k})},\pi\cos{(\theta_{k})},2\pi\Delta f\tau_{k}\big]^T$ and $\hat{\boldsymbol{\omega}}_{l}^{(\rm p)}=\big[\frac{2\pi\tilde{n}_{1,l}}{O_{\rm h}N_{\rm h}},\frac{2\pi\tilde{n}_{2,l}}{O_{\rm v}N_{\rm v}},\frac{2\pi\tilde{n}_{3,l}}{O_{\rm d}K_{\rm m}}\big]^T$. Then, the weak partition $\{\mathcal{K}_{1}^*,\ldots,\mathcal{K}_{\hat{L}}^*\}$ is yielded based on mapping $k\mapsto\mathop{\arg\min}_{l}\{D(\boldsymbol{\omega}_{k},\hat{\boldsymbol{\omega}}_{l}^{(\rm p)})\}$ and UB-NPAE in \eqref{equ: UB-NPAE-1} can be calculated. Since $\text{UB-NPAE}\geq\text{NPAE}$ is intuitively held, it can also indicate the accurate physical-association when UB-NPAE approaches zero. 

\section{Proof of Proposition~\ref{prop: transformed label}}
\label{appdix: transformation proof}
Based on \eqref{equ: aligned feature}, the aligned path response $\widetilde{\mathbf{A}}_{l}$ can be represented by
\begin{equation}
    \label{equ: aligned path feature}
    \widetilde{\mathbf{A}}_{l}\!=\!\sum_{k\in\mathcal{I}^{(\rm m)}_{l}}\hat{\alpha}_{k}^{(\rm m)}e^{-{\rm j}2\pi f_{1}^{(\rm m)}\hat{\tau}_{k}^{(\rm m)}}\mathbf{a}(\tilde{\varphi}_{k}^{(\rm m)}, \tilde{\theta}_{k}^{(\rm m)})\mathbf{b}_{\rm m}(\tilde{\tau}_{k}^{(\rm m)}),
\end{equation}
where the parameters $(\tilde{\varphi}_{k}^{(\rm m)},\tilde{\theta}_{k}^{(\rm m)},\tilde{\tau}_{k}^{(\rm m)})$ are yielded by shifting $(\hat{\varphi}_{k}^{(\rm m)},\hat{\theta}_{k}^{(\rm m)},\hat{\tau}_{k}^{(\rm m)})$ as
\begin{equation}
\label{equ: shifted parameters}
\begin{aligned}
&\sin({\tilde{\varphi}_{k}}^{(\rm m)})\sin(\tilde{\theta}_{k}^{(\rm m)})\!=\!\sin({\hat{\varphi}_{k}^{(\rm m)}})\sin(\hat{\theta}_{k}^{(\rm m)})\!-\!\frac{2\tilde{n}_{1,l}}{O_{\rm h}N_{\rm h}},\\
&\cos(\tilde{\theta}_{k}^{(\rm m)})\!=\!\cos(\hat{\theta}_{k}^{(\rm m)})\!-\!\frac{2\tilde{n}_{2,l}}{O_{\rm v}N_{\rm v}},\tilde{\tau}_{k}^{(\rm m)}\!=\!\hat{\tau}_{k}^{(\rm m)}\!-\!\frac{\tilde{n}_{3,l}}{O_{\rm d}K_{\rm m}\Delta f}\\
\end{aligned}
\end{equation}
for all $k\in\mathcal{I}_{l}^{(\rm m)}$. Based on the spatial reciprocity of measured and target frequency band, parameters $(\hat{\varphi}_{k}^{(\rm e)},\hat{\theta}_{k}^{(\rm e)},\hat{\tau}_{k}^{(\rm e)}), k\in\mathcal{I}_{l}^{(\rm e)}$ in $\widehat{\mathbf{B}}_{l}$ should apply the same shifts as in \eqref{equ: shifted parameters}. Meanwhile, phase rotation term $\beta(\tilde{n}_{3,l})$ in \eqref{equ: beta} should also be considered due to the delay shift $-\frac{\tilde{n}_{3,l}}{O_{\rm d}K_{\rm m}\Delta f}$, which eventually yields the co-transformed response $\widetilde{\mathbf{B}}_{l}$ in \eqref{equ: transformed label}. 

\section{Proof of Theorem~\ref{theo: bound original}}
\label{appdix: bound 1}
Under proper path extraction design, the channel in the target frequency band can be approximately decomposed as 
\begin{equation}
    \label{equ: approx}
    \mathbf{H}^{(\rm e)}\approx\sum_{l=1}^{\hat{L}}\widehat{\mathbf{B}}_{l}=\sum_{l=1}^{\hat{L}}\text{conj}(\mathbf{V}_{l})\odot\widetilde{\mathbf{B}}_{l}.
\end{equation}
Thus, $\mathcal{J}(\eta^{*})$ can be upper bounded by 
\begin{subequations}
    \begin{align}
   &\mathcal{J}(\eta^{*})\!\approx\!\mathbb{E}\Big\{\!\big\Vert\sum_{l}\text{conj}(\mathbf{V}_{l})\!\odot\!\Big(\widetilde{\mathbf{B}}_{l}\!-\!\eta^{*}(\widetilde{\mathbf{A}}_{l})\Big)\big\Vert_{F}^{2}\!\Big\},\\
   &\leq\mathbb{E}\Big\{\hat{L}\sum_{l}\big\Vert\text{conj}(\mathbf{V}_{l})\!\odot\!\Big(\widetilde{\mathbf{B}}_{l}-\eta^{*}(\widetilde{\mathbf{A}}_{l})\Big)\big\Vert_{F}^{2}\Big\},\label{subeq: basic inequality}\\
   &=\mathbb{E}\Big\{\hat{L}\sum_{l}\big\Vert\widetilde{\mathbf{B}}_{l}-\eta^{*}(\widetilde{\mathbf{A}}_{l})\big\Vert_{F}^{2}\Big\},\label{subeq: constant norm}\\
   &\approx\mathbb{E}_{{Q}_{\hat{L}}}\{\hat{L}^{2}\}\mathbb{E}_{Q^{(\rm pa)}}\Big\{\big\Vert\widetilde{\mathbf{B}}_{l}-\eta^{*}(\widetilde{\mathbf{A}}_{l})\big\Vert_{F}^{2}\Big\}\label{subeq: unconditional},
   \end{align}
\end{subequations}
where \eqref{subeq: basic inequality} is held due to the inequality $\Vert\sum_{i=1}^{n}\mathbf{X}_{i}\Vert_{F}^{2}\leq n\sum_{i=1}^{n}\Vert\mathbf{X}_{i}\Vert_{F}^2$,  \eqref{subeq: constant norm} is held since all the elements in $\mathbf{V}_{l}$ have unit modulus, and \eqref{subeq: unconditional} is held due to \textit{Assumption~\ref{asup: unconditional}}. 

Denote $\mathcal{L}_{Q^{(\rm pa)}}(\eta^{*})=\mathbb{E}_{Q^{(\rm pa)}}\big\{\Vert\widetilde{\mathbf{B}}_{l}-\eta^{*}(\widetilde{\mathbf{A}}_{l})\Vert_{F}^{2}\big\}$. Then, upper bound of $\mathcal{L}_{Q^{(\rm pa)}}(\eta^{*})$ is derived as below. For the DL extrapolator $\eta\in\mathcal{F}$, the MSE loss for input $\widetilde{\mathbf{A}}$ can be expressed as $g_{\eta}(\widetilde{\mathbf{A}})=\Vert\widetilde{\mathbf{B}}-\eta(\widetilde{\mathbf{A}})\Vert_{F}^{2}=\Vert\psi(\widetilde{\mathbf{A}})-\eta(\widetilde{\mathbf{A}})\Vert_{F}^{2}$. Then, we first prove that $g_{\eta}(\widetilde{\mathbf{A}})$ is Lipschitz continuous. For all $\widetilde{\mathbf{A}}_{1},\widetilde{\mathbf{A}}_{2}\in\text{supp}(P_{\widetilde{\mathbf{A}}}^{(\rm pa)})\cup\text{supp}(Q_{\widetilde{\mathbf{A}}}^{(\rm pa)})$, it can be derived that $|g_{\eta}(\widetilde{\mathbf{A}}_{1})-g_{\eta}(\widetilde{\mathbf{A}}_{2})|=g_{1}\times g_{2}$, where
\begin{equation}
    \label{equ: lip}
    \begin{aligned}
    &g_{1}=\Vert\psi(\widetilde{\mathbf{A}}_{1})-\eta(\widetilde{\mathbf{A}}_{1})\Vert_{F}+\Vert\psi(\widetilde{\mathbf{A}}_{2})-\eta(\widetilde{\mathbf{A}}_{2})\Vert_{F}\\
    &g_{2}=\big|\Vert\psi(\widetilde{\mathbf{A}}_{1})-\eta(\widetilde{\mathbf{A}}_{1})\Vert_{F}-\Vert\psi(\widetilde{\mathbf{A}}_{2})-\eta(\widetilde{\mathbf{A}}_{2})\Vert_{F}\big|.
    \end{aligned}
\end{equation}
Based on triangle inequality, bounded support in \textit{Assumption~\ref{asup: boundded support}} and Lipschitz continuity in \textit{Assumption~\ref{asup: Lip}}, $g_{1}$ can be upper bounded by 
\begin{equation}
    \begin{aligned}
    \label{equ: g1}
    g_{1}&\leq\Vert\psi(\widetilde{\mathbf{A}}_{1})\Vert_{F}\!+\!\Vert\eta(\widetilde{\mathbf{A}}_{1})\Vert_{F}\!+\!\Vert\psi(\widetilde{\mathbf{A}}_{2})\Vert_{F}\!+\!\Vert\eta(\widetilde{\mathbf{A}}_{2})\Vert_{F}\\
    &\leq 4R_{1}R_{2}.
    \end{aligned}
\end{equation}
Similarly, $g_{2}$ can be upper bounded by
\begin{equation}
    \begin{aligned}
    \label{equ: g2}
    g_{2}&\leq\Vert\psi(\widetilde{\mathbf{A}}_{1})-\eta(\widetilde{\mathbf{A}}_{1})-\psi(\widetilde{\mathbf{A}}_{2})+\eta(\widetilde{\mathbf{A}}_{2})\Vert_{F}\\
    &\leq\Vert\psi(\widetilde{\mathbf{A}}_{1})-\psi(\widetilde{\mathbf{A}}_{2})\Vert_{F}+\Vert\eta(\widetilde{\mathbf{A}}_{1})-\eta(\widetilde{\mathbf{A}}_{2})\Vert_{F}\\
    &\leq 2R_{2}\Vert\widetilde{\mathbf{A}}_{1}-\widetilde{\mathbf{A}}_{2}\Vert_{F}
    \end{aligned}
\end{equation}
Combining \eqref{equ: g1} and \eqref{equ: g2}, it can be derived that 
\begin{equation}
    \label{equ: Lipschitz}
    \begin{aligned}
    |g_{\eta}(\widetilde{\mathbf{A}}_{1})-g_{\eta}(\widetilde{\mathbf{A}}_{2})|&\leq 8 R_{1}R_{2}^{2}\Vert\widetilde{\mathbf{A}}_{1}-\widetilde{\mathbf{A}}_{2}\Vert_{F}\\
    &\triangleq C\Vert\widetilde{\mathbf{A}}_{1}-\widetilde{\mathbf{A}}_{2}\Vert_{F}.
    \end{aligned}
\end{equation}
Thus, $g_{\eta}(\widetilde{\mathbf{A}})$ is $C$-Lipschitz continuous for any possible DL extrapolator $\eta\in\mathcal{F}$. 

Subsequently, we first decompose $\mathcal{L}_{Q^{(\rm pa)}}(\eta^{*})$ into three components, namely, minimal training loss, intra-environment generalization gap, and inter-environment generalization gap. To simplify the expression, we omit the superscripts $(\rm pa)$ in $Q^{(\rm pa)}, \mathcal{D}^{(\rm pa)}, P^{(\rm pa)}$ in the following derivations. Then, it can be derived that
\begin{equation}
    \label{equ: decompose}
    \begin{aligned}
    &\mathcal{L}_{Q}(\eta^{*})\!=\!\mathcal{L}_{\mathcal{D}}(\eta^{*})\!+\!\mathcal{L}_{P}(\eta^{*})\!-\!\mathcal{L}_{\mathcal{D}}(\eta^{*})\!+\!\mathcal{L}_{Q}(\eta^{*})\!-\!\mathcal{L}_{P}(\eta^{*})\\
    &\!\leq\!\mathcal{L}_{\mathcal{D}}(\eta^{*})\!+\!\sup_{\eta\in\mathcal{F}}|\mathcal{L}_{\mathcal{D}}(\eta)\!-\!\mathcal{L}_{P}(\eta)|\!+\!\sup_{\eta\in\mathcal{F}}|\mathcal{L}_{P}(\eta)\!-\!\mathcal{L}_{Q}(\eta)|.
    \end{aligned}
\end{equation}
Based on Kantorovich-Rubinstein duality \cite[Theorem 5.10]{villani2009optimal}, Wasserstein-1 distance $W_{1}(P_{\widetilde{\mathbf{A}}},Q_{\widetilde{\mathbf{A}}})$ for marginal distribution $P_{\widetilde{\mathbf{A}}}$ and $Q_{\widetilde{\mathbf{A}}}$ can be equivalently reformulated as 
\begin{equation}
    \label{equ: wasserstein}
    W_{1}(P_{\widetilde{\mathbf{A}}},Q_{\widetilde{\mathbf{A}}})=\sup_{\Vert g\Vert_{L}\leq1}\big|\mathbb{E}_{P_{\widetilde{\mathbf{A}}}}\{g(\widetilde{\mathbf{A}})\}-\mathbb{E}_{Q_{\widetilde{\mathbf{A}}}}\{g(\widetilde{\mathbf{A}})\}\big|,
\end{equation}
where $\Vert g \Vert_{L}=\sup_{\widetilde{\mathbf{A}}_{1},\widetilde{\mathbf{A}}_{2}}\frac{|g(\widetilde{\mathbf{A}}_{1})-g(\widetilde{\mathbf{A}}_{2})|}{\Vert\widetilde{\mathbf{A}}_{1}-\widetilde{\mathbf{A}}_{2}\Vert_{F}}$ denotes the Lipschitz semi-norm. Based on the definition and Lipschitz continuity of $g_{\eta}$, the inter-environment generalization gap $\Delta_{\rm env}=\sup_{\eta\in\mathcal{F}}|\mathcal{L}_{P}(\eta)-\mathcal{L}_{Q}(\eta)|$ can be bounded by 
\begin{equation}
    \label{equ: gap}
    \begin{aligned}
    \Delta_{\rm env}&\leq \sup_{\Vert g\Vert_{L}\leq C}\big|\mathbb{E}_{P_{\widetilde{\mathbf{A}}}}\{g(\widetilde{\mathbf{A}})\}-\mathbb{E}_{Q_{\widetilde{\mathbf{A}}}}\{g(\widetilde{\mathbf{A}})\}\big|\\
    &=C W_{1}(P_{\widetilde{\mathbf{A}}},Q_{\widetilde{\mathbf{A}}}).
    \end{aligned}
\end{equation}
By combining \eqref{equ: decompose}, \eqref{equ: gap} and \eqref{subeq: unconditional}, upper bound of $\mathcal{J}(\eta^{*})$ in \eqref{equ: upper bound JQtilde} can be proved. This completes the proof. 

\ifCLASSOPTIONcaptionsoff
  \newpage
\fi

\bibliographystyle{IEEEtran}
\bibliography{ref}

\end{document}